%% file: aroma_v2.tex
\documentclass[acmsmall,screen]{acmart}

\setcopyright{rightsretained}
\acmPrice{}
\acmDOI{10.1145/3360578}
\acmYear{2019}
\copyrightyear{2019}
\acmJournal{PACMPL}
\acmVolume{3}
\acmNumber{OOPSLA}
\acmArticle{152}
\acmMonth{10}

\usepackage{booktabs}   %% For formal tables:
\usepackage{subcaption} %% For complex figures with subfigures/subcaptions
\definecolor{dycolor}{rgb}{0.5,0.7,0.9}
\definecolor{flcolor}{rgb}{0.05,0.4,0.1}
\definecolor{kscolor}{rgb}{0.9,0.1,0.1}
\definecolor{sccolor}{rgb}{0.9,0.7,0.1}
\newcommand\fl[1]{}
\newcommand\cb[1]{}
\newcommand\dy[1]{}
\newcommand\ksen[1]{}
\newcommand\sch[1]{}
\newcommand\todo[1]{}
\newcommand\review[1]{}

\usepackage{algorithm}
\usepackage{algpseudocode}
\usepackage{amsthm}
\usepackage{array}
\usepackage{balance}
\usepackage{booktabs}
\usepackage{caption}
\usepackage{color}
\usepackage{enumitem}
\usepackage{listings}
\usepackage{perpage}
\usepackage{soul}
\usepackage{threeparttable}
\usepackage{wrapfig}

\usepackage[T1]{fontenc}

\renewcommand{\thefootnote}{\fnsymbol{footnote}}
\MakePerPage{footnote}

\definecolor{hlgray}{gray}{0.8}
\sethlcolor{hlgray}
\newcommand{\explanation}[1]{
    \parbox{0.45\textwidth}{
        \rule{0.4\textwidth}{0.1pt}
        \vspace*{0.5em} \\
        {#1}
    }
}

\newtheorem{definition}{Definition}

\def\aroma{\textsc{Aroma}}

\begin{document}

% %% Title information
\title{Aroma: Code Recommendation via Structural Code Search}
% \title[Short Title]{Full Title}         %% [Short Title] is optional;
%                                         %% when present, will be used in
%                                         %% header instead of Full Title.
% \titlenote{with title note}             %% \titlenote is optional;
%                                         %% can be repeated if necessary;
%                                         %% contents suppressed with 'anonymous'
% \subtitle{Subtitle}                     %% \subtitle is optional
% \subtitlenote{with subtitle note}       %% \subtitlenote is optional;
%                                         %% can be repeated if necessary;
%                                         %% contents suppressed with 'anonymous'

% %% Author information
% %% Contents and number of authors suppressed with 'anonymous'.
% %% Each author should be introduced by \author, followed by
% %% \authornote (optional), \orcid (optional), \affiliation, and
% %% \email.
% %% An author may have multiple affiliations and/or emails; repeat the
% %% appropriate command.
% %% Many elements are not rendered, but should be provided for metadata
% %% extraction tools.

\author{Sifei Luan}
\affiliation{
  \institution{Facebook}
  \city{Menlo Park}
  \state{CA}
  \country{USA}
}
\email{lsf@fb.com}

\author{Di Yang}
\authornote{This work was done in part while this author was an intern at Facebook.}
\affiliation{
  \institution{University of California, Irvine}
  \city{Irvine}
  \state{CA}
  \country{USA}
}
\email{diy4@uci.edu}

\author{Celeste Barnaby}
\affiliation{
  \institution{Facebook}
  \city{Menlo Park}
  \state{CA}
  \country{USA}
}
\email{celestebarnaby@fb.com}

\author{Koushik Sen}
\authornote{This work was done in part while this author was a visiting scientist at Facebook.}
\affiliation{
  \institution{University of California, Berkeley}
  \city{Berkeley}
  \state{CA}
  \country{USA}
}
\email{ksen@cs.berkeley.edu}

\author{Satish Chandra}
\affiliation{
  \institution{Facebook}
  \city{Menlo Park}
  \state{CA}
  \country{USA}
}
\email{satch@fb.com}

% Suppress the Authors addresses footnote to avoid page breaks
\makeatletter
\let\@authorsaddresses\@empty
\makeatother

%% Abstract
%% Note: \begin{abstract}...\end{abstract} environment must come
%% before \maketitle command
\input{abstract}

%% 2012 ACM Computing Classification System (CSS) concepts
%% Generate at 'http://dl.acm.org/ccs/ccs.cfm'.

%% End of generated code

%% Keywords
%% comma separated list

%% \maketitle
%% Note: \maketitle command must come after title commands, author
%% commands, abstract environment, Computing Classification System
%% environment and commands, and keywords command.
\maketitle

%------------------------------------------------------------

\sloppy

% Use numbers for footnote in main body
\renewcommand{\thefootnote}{\arabic{footnote}}

\input{intro}
\input{similar-code-study}
\input{algorithm}
\input{evaluation}

\input{microbenchmark}

\input{userstudy}
\input{related}

\input{conclusion}

\bibliography{aroma}

\end{document}

%% file: abstract.tex
\begin{abstract}
% \ksen{Fix the abstract because we have new sections: implementation for 4 languages, developer study, etc.} \cb{I added the additional sections}
Programmers often write code that has similarity to existing code written somewhere. A tool that could help programmers to search such similar code would be immensely useful. Such a tool could help programmers to extend partially written code snippets to completely implement necessary functionality, help to discover extensions to the partial code which are commonly included by other programmers, help to cross-check against similar code written by other programmers, or help to add extra code which would fix common mistakes and errors. We propose Aroma, a tool and technique for code recommendation via structural code search. Aroma indexes a huge code corpus including thousands of open-source projects, takes a partial code snippet as input, searches the corpus for method bodies containing the partial code snippet, and clusters and intersects the results of the search to recommend a small set of succinct code snippets which both contain the query snippet and appear as part of several methods in the corpus. We evaluated Aroma on 2000 randomly selected queries created from the corpus, as well as 64 queries derived from code snippets obtained from Stack Overflow, a popular website for discussing code. We implemented Aroma for 4 different languages, and developed an IDE plugin for Aroma. Furthermore, we conducted a study where we asked 12 programmers to complete programming tasks using Aroma, and collected their feedback. Our results indicate that Aroma is capable of retrieving and recommending relevant code snippets efficiently.
% \ksen{All latex errors must be fixed.}
\end{abstract}

% \begin{IEEEkeywords}
% code recommendation, structural code search
% \end{IEEEkeywords}

%% file: intro.tex
\section{Introduction}

% \ksen{The intro on the first page seems to have two columns which must be avoided.}
% Suppose a programmer wants to write code to decode a bitmap without incurring a high memory overhead.  The programmer is familiar with the libraries necessary to write the code, but they are not quite sure how to write the code completely with proper error handling and suitable configurations.  They write the partial snippet shown in Table~\ref{tab:intro-examples}, row A, left column.  The programmer now wants to know how others have implemented this functionality fully and correctly in related projects.  Specifically, they want to know what is the customary way to extend the code so that proper setup is done, common errors are handled, and appropriate library methods are called.  It would be nice if a tool could return the code in Table~\ref{tab:intro-examples}, row A, right column, which shows how to handle potential out-of-memory errors; it also shows how the configurable options for decoding the bitmap are commonly set.  We call this the \emph{code recommendation problem.}

Suppose an Android programmer wants to write code to decode a bitmap.  The programmer is familiar with the libraries necessary to write the code, but they are not quite sure how to write the code completely with proper error handling and suitable configurations.  They write the code snippet shown in Listing~\ref{lst:intro-bitmap-query} as a first attempt.  The programmer now wants to know how others have implemented this functionality fully and correctly in related projects.  Specifically, they want to know what is the customary way to extend the code so that proper setup is done, common errors are handled, and appropriate library methods are called.  It would be nice if a tool could return a few code snippets shown in Listings~\ref{lst:intro-bitmap-rec-1}, \ref{lst:intro-bitmap-rec-2}, which demonstrate how to configure the decoder to use less memory, and how to handle potential runtime exceptions, respectively.  We call this the \emph{code recommendation problem.}  \review{The examples below are new, but they are all open-source code, no FB-internal code.}

\captionsetup[lstlisting]{singlelinecheck=false}
\begin{lstlisting}[label={lst:intro-bitmap-query},caption={Suppose an Android programmer writes this code to decode a bitmap.}]
InputStream input = manager.open(fileName);
Bitmap image = BitmapFactory.decodeStream(input);
\end{lstlisting}
\captionsetup[lstlisting]{singlelinecheck=true}

\begin{lstlisting}[label={lst:intro-bitmap-rec-1},caption={A recommended code snippet that shows how to configure the decoder to use less memory. Recommended lines are highlighted.\protect\footnotemark}]
(*@\hl{final BitmapFactory.Options options = new BitmapFactory.Options();}@*)
(*@\hl{options.inSampleSize = 2;}@*)
Bitmap bmp = BitmapFactory.decodeStream(is, null, options);
\end{lstlisting}
\footnotetext{Adapted from \url{https://github.com/zom/Zom-Android/blob/master/app/src/main/java/org/awesomeapp/messenger/ui/stickers/StickerGridAdapter.java\#L67}. Accessed in August 2018.}

% \addtocounter{endnote}{-1}
\begin{lstlisting}[label={lst:intro-bitmap-rec-2},caption={Another recommended code snippet that shows how to properly close the input stream and handle any potential \texttt{IOException}. Recommended lines are highlighted.\protect\footnotemark}]
(*@\hl{try \{}@*)
    InputStream is = am.open(fileName);
    image = BitmapFactory.decodeStream(is);
    (*@\hl{is.close();}@*)
(*@\hl{\} catch (IOException e) \{}@*)
    (*@\hl{// ...}@*)
(*@\hl{\}}@*)
\end{lstlisting}
\footnotetext{Adapted from \url{https://github.com/yuyuyu123/ZCommon/blob/master/zcommon/src/main/java/com/cc/android/zcommon/utils/android/AssetUtils.java\#L37}. Accessed in August 2018.}

% Modern day developers work with complex software frameworks which contain a huge amount of APIs. A key challenge is writing code that is safe, correct, and idiomatic. The availability of large amount of existing code provides an opportunity: If a developer is presented with code snippets similar to what they are writing, they could learn the idiomatic usage and the best practices from these real-world examples. Furthermore, if a system could automatically cluster the similar code to generate succinct code recommendations containing the popular patterns, it could be immensely useful to the developer because they can use these recommendations for completing the necessary functionality, discovering common patterns to extend a functionality, or cross-checking to avoid common mistakes and errors.

There are a few existing techniques which could potentially be used to get code recommendations.  For example, code-to-code search tools~\cite{Kim:2018,krugle} could retrieve relevant code snippets from a corpus using a partial code snippet as query.  However, such code-to-code search tools return lots of relevant code snippets without removing or aggregating similar-looking ones.  Moreover, such tools do not make any effort to carve out common and concise code snippets from similar-looking retrieved code snippets. Pattern-based code completion tools~\cite{Nguyen:2009, Mover:2018, Nguyen:2012} mine common API usage patterns from a large corpus and use those patterns to recommend code completion for partially written programs as long as the partial program matches a prefix of a mined pattern.  Such tools work well for the mined patterns; however, they cannot recommend any code outside the mined patterns---the number of mined patterns are usually limited to a few hundreds.  We emphasize that the meaning of the phrase ``code recommendation'' in \aroma{} is different from the term ``API code recommendation''~\cite{Nguyen:2016:APIRec, Nguyen:2016:bytecode}. The latter is a recommendation engine for the next API method to invoke given a code change, whereas \aroma{} aims to recommend code snippets, as shown in Listings~\ref{lst:intro-bitmap-rec-1}, \ref{lst:intro-bitmap-rec-2}, for programmers to learn common usages and integrate those usages with their own code.  \aroma{}'s recommendations contain more syntactic variety than just API usages; for instance, the recommended code snippet in Listing~\ref{lst:intro-bitmap-rec-2} includes a \texttt{try-catch} block, and Example B in Table~\ref{tab:intro-examples} recommends adding an \texttt{if} statement that modifies a variable. Code clone detectors~\cite{Sajnani:2016, Cordy:2011, Jiang:2007, Kamiya:2002} are another set of techniques that could potentially be used to retrieve recommended code snippets.  However, code clone detection tools usually retrieve code snippets that are almost identical to a query snippet.  Such retrieved code snippets may not always contain extra code which could be used to extend the query snippet.

We propose \aroma{}, a code recommendation engine. Given a code snippet as input query and a large corpus of code containing millions of methods, \aroma{} returns a set of recommended code snippets such that each recommended code snippet:
\begin{itemize}
	\item  contains the query snippet approximately, and
	\item  is contained approximately in a non-empty set of method bodies in the  corpus.
\end{itemize}
Furthermore, \aroma{} ensures that any two recommended code snippets are not quite similar to each other.

%We emphasize that the \textbf{code recommendation} \ksen{unnecessary bold fonts should be avoided.} capability of \aroma{} is different from the term ``API code recommendation''~\cite{Nguyen:2016:APIRec, Nguyen:2016:bytecode}. The latter is a recommendation engine for the next API method to invoke given a code change, where \aroma{} aims to provide various recommended code snippets, as shown in Listings~\ref{lst:intro-bitmap-rec-1}, \ref{lst:intro-bitmap-rec-2}, for programmers to learn common usages and integrate with their own code. \fl{Added this paragraph to disambiguate from APIRec.} \ksen{This para looks good in isolation, but kind of looks awkward in the flow of the intro.  It sounds like we are trying hard to compare against API recommendation work. I am not sure how to integrate it coherently with the rest of the intro.  Any ideas?}

\aroma{} works by first indexing the given corpus of code.  Then \aroma{} searches for a small set (e.g. 1000) of method bodies which contain the query code snippet \textit{approximately}.
%\ksen{Commenting this: We have designed this step to finish in less than a second on a server machine.}
%CPU \review{This information about hardware is new.}.
% For this search step, we cannot use a clone detection tool because clone detection tools are good at retrieving code snippets that are almost identical to the query snippet---\aroma{}'s first step needs to retrieve code snippets that \emph{contain} the query snippet in addition to extra code.
A challenge in designing this search step is that a query snippet, unlike a natural language query, has structure, which should be taken into account while searching for code.  Once \aroma{} has retrieved a small set of code snippets which approximately contain the query snippet, \aroma{} prunes the retrieved snippets so that the resulting pruned snippets become similar to the query snippet.  It then ranks the retrieved code snippets based on the similarity of the pruned snippets to the query snippet.  This step helps to rank the retrieved snippets based on how well they contain the query snippet.  The step is precise, but is relatively expensive; however, the step is only performed on a small set of code snippets, making it efficient in practice.  After ranking the retrieved code snippets, \aroma{} clusters the snippets so that similar snippets fall under the same cluster.  \aroma{} then intersects the snippets in each cluster to carve out a maximal code snippet which is common to all the snippets in the cluster and which contains the query snippet.  The set of intersected code snippets are then returned as recommended code snippets.  Figure~\ref{fig:aroma_model} shows an outline of the algorithm.  For the query shown in Listing~\ref{lst:intro-bitmap-query}, \aroma{} recommends the code snippets shown in Listings~\ref{lst:intro-bitmap-rec-1}, \ref{lst:intro-bitmap-rec-2}. The right column of Table~\ref{tab:intro-examples} shows more examples of code snippets recommended by \aroma{} for the code queries shown on the left column of the table.

\input{intro-table}

To our best knowledge, \aroma{} is the first tool which could recommend relevant code snippets given a query code snippet.  The advantages of \aroma{} are the following:
\begin{itemize}
\item A code snippet recommended by \aroma{} does not simply come from a single method body, but is generated from several similar-looking code snippets via intersection.  This increases the likelihood that \aroma{}'s recommendation is idiomatic rather than one-off.
\item \aroma{} does not require mining common coding patterns or idioms ahead of time.  Therefore, \aroma{} is not limited to a set of mined patterns---it can retrieve new and interesting code snippets on-the-fly.
\item \aroma{} is fast enough to use in real time.
%Although \aroma{}'s pruning and clustering algorithms are time-consuming and one cannot afford to run these algorithms for millions of methods,
A key innovation in \aroma{} is that it first retrieves a small set of snippets based on approximate search, and then performs the heavy-duty pruning and clustering operations on this set.  This enables \aroma{} to create recommended code snippets on a given query from a large corpus containing millions of methods within a couple of seconds on a multi-core server machine.
\item
%\ksen{the following two sentences are similar to what we mention in the previous bullet point: \aroma{} is scalable to codebases that contain millions of methods.  We evaluate \aroma{} on a open-source code corpus containing 2.4 million methods to match the scale of codebases in large companies.}
\aroma{} is easy to deploy for different programming languages because its core algorithm works on generic parse trees. We have implemented \aroma{} for Hack, Java, JavaScript and Python.
%\ksen{We do not use the term "generic syntax trees" in the algorithms section.  We should be precise while using terminologies.}
%\fl{Added a bullet point here in response to Michael's comment: A major contribution is the scalability of the engine. The paper should make this more explicit, both in the approach and the evaluation sections.} \ksen{How is this different from the previous bullet point which is also mentions scalability.}
\item Although we developed \aroma{} for the purpose of code recommendation, it could be used to also perform efficient and precise code-to-code structural search.
%Such search could help programmers to check if they have written their code in an idiomatic way or not.
\end{itemize}

We have implemented \aroma{} in C++ for four programming languages: Hack~\cite{hacklang}, Java, JavaScript and Python. We have also implemented IDE plugins for all of these four languages.  We report our experimental evaluation of \aroma{} for the Java programming language.  We have used \aroma{} to index 5,417 GitHub Java Android projects. We performed our experiments for Android Java because we initially developed \aroma{} for Android based on internal developers' need. We evaluated \aroma{} using code snippets obtained from Stack Overflow.  We manually analyzed and categorized the recommendations into several representative categories.  We also evaluated \aroma{} recommendations on 50 partial code snippets, where we found that \aroma{} can recommend the exact code snippets for 37 queries, and in the remaining 13 cases \aroma{} recommends alternative recommendations that are still useful.  On average, \aroma{} takes 1.6 seconds to create recommendations for a query code snippet on a 24-core CPU.  In our large-scale automated evaluation, we used a micro-benchmarking suite containing artificially created query snippets to evaluate the effectiveness of various design choices in \aroma{}. Finally, we conducted a user study of \aroma{} by observing 12 Hack programmers interacting with the IDE on 4 short programming tasks, and found that \aroma{} is a helpful addition to the existing coding assistant tools.

The rest of the paper is organized as follows: Section~\ref{sec:similar-code-study} presents a case study that reveals the opportunity for a code recommendation tool like \aroma{}. In Section~\ref{sec:algorithm}, we describe the algorithm \aroma{} uses to create code recommendations. In Section~\ref{sec:evaluation} we manually assess how useful \aroma{} code recommendations are. Since code search is a key component of creating recommendations, in Section~\ref{sec:microbenchmark} we measure the search recall of \aroma{} and compare it with other techniques. Section~\ref{sec:more-languages} introduces the real-world deployment of \aroma{}. In Section~\ref{sec:userstudy}, we report the initial developer experience with using the \aroma{} tool. Section~\ref{sec:related} presents related work. Finally, Section~\ref{sec:conclusion} concludes the paper.

%We further evaluated the precision of the prune and rerank phase and justified the design choices of \aroma{} using a micro-benchmark dataset of randomly selected queries created from the corpus.

%The rest of the paper is organized as follows: In Section \ref{sec:algorithm}, we describe the algorithm \aroma{} uses for generating recommendations. Section \ref{sec:microbenchmark} presents the micro-benchmark results where different design choices are evaluated. In Section \ref{sec:evaluation} we evaluate the quality of Aroma recommendations results. Finally, Section \ref{sec:related} presents related work.\fl{I plan to move micro-benchmark before evaluation. They are currently reversed.}

% Aroma is a code recommendation engine.  Given a code snippet, called the query snippet,  and a big code corpus, it retrieves code snippets from the corpus which contain a significant portion of the query snippet.  Such recommended code snippets could help programmers in the following ways:

% \begin{itemize}
% \item Recommended code snippets could help  to understand how the query snippet could be extended or completed..  Programmers could copy-paste a recommended code snippet and integrate it with the query code snippet.
% \item Recommended code snippets could help to programmers to identify idiomatic use of the query code and could suggest modifications to the query code so that the query code becomes more idiomatic.
% \item Aroma could help programmers to perform structural search over the code corpus.
% \end{itemize}

%% file: intro-table.tex
% Set listing style for table
% \ksen{Move Table 2 to page 2.}
\lstset{
	frame=none,
    aboveskip=0pt,
    belowskip=0pt,
    basicstyle=\linespread{0.5}\tiny\ttfamily,
}
\begin{table*}\tiny
\begin{threeparttable}
\caption{\aroma{} code recommendation examples}
\label{tab:intro-examples}
% \ksen{For all query and recommended code snippet pairs shown in various listings, we need the number of snippets that were intersected to obtain the recommended code.}

\setlength{\tabcolsep}{0.01\textwidth}
\begin{tabular}{@{}p{0.49\textwidth}p{0.49\textwidth}@{}}
\toprule
Query Code Snippet & \aroma{} Code Recommendation with Extra Lines Highlighted \\
\midrule

\begin{lstlisting}
TextView textView = (TextView) view.findViewById(R.id.textview);
SpannableString content = new SpannableString("Content");
content.setSpan(new UnderlineSpan(), 0, content.length(), 0);
textView.setText(content);
\end{lstlisting}
\vspace*{1em}
\explanation{
    \emph{Example A: Configuring Objects.}
    \begin{itemize}[leftmargin=*]
        \item This code snippet adds underline to a piece of text.\tnote{1} % ~\cite{SO:underline}
        \item The recommended code suggests adding a callback handler to pop up a dialog once the underlined text is touched upon.
        \item \textbf{Intersected from a cluster of 2 methods}.\tnote{2} % ~\cite{GH:intro5, GH:intro6}
    \end{itemize}
}
% https://github.com/tonyvu2014/android-shoppingcart/blob/master/demo/src/main/java/com/android/tonyvu/sc/demo/ProductActivity.java#L58
% https://github.com/tonyvu2014/android-shoppingcart/blob/master/demo/src/main/java/com/android/tonyvu/sc/demo/MainActivity.java#L23
&
\begin{lstlisting}
TextView licenseView = (TextView) findViewById(R.id.library_license_link);
SpannableString underlinedLicenseLink = new SpannableString(
    getString(R.string.library_license_link));
underlinedLicenseLink.setSpan(new UnderlineSpan(), 0, underlinedLicenseLink.length(), 0);
licenseView.setText(underlinedLicenseLink);
(*@\hl{licenseView.setOnClickListener(v -> \{}@*)
    (*@\hl{FragmentManager fm = getSupportFragmentManager();}@*)
    (*@\hl{LibraryLicenseDialog libraryLicenseDlg = new LibraryLicenseDialog();}@*)
    (*@\hl{libraryLicenseDlg.show(fm, "fragment\_license"); \});}@*)
\end{lstlisting}
\\

\bottomrule

\begin{lstlisting}
Bitmap bitmap = BitmapFactory.decodeResource(getResources(), R.drawable.image);
\end{lstlisting}
\vspace*{0.5em}
\explanation{
    \emph{Example B: Post-Processing.}
    \begin{itemize}[leftmargin=*]
        \item This code snippet decodes a bitmap.\tnote{3} % ~\cite{SO:bitmap}
        \item The recommended code suggests applying Gaussian blur on the decoded image, a customary effect to be applied.
        \item \textbf{Intersected from a cluster of 4 methods}.\tnote{4} % ~\cite{GH:intro7,GH:intro8,GH:intro9,GH:intro10}
    \end{itemize}
}
% https://github.com/TonnyL/GaussianBlur/blob/master/app/src/main/java/io/github/marktony/gaussianblur/MainActivity.java#L58
% https://github.com/TonnyL/GaussianBlur/blob/master/app/src/main/java/io/github/marktony/gaussianblur/MainActivity.java#L28
% https://github.com/irccloud/android/blob/master/src/com/irccloud/android/activity/ImageViewerActivity.java#L149
% https://github.com/REBOOTERS/My-MVP/blob/master/app/src/main/java/huyifei/mymvp/GlideActivity.java#L34
&
\begin{lstlisting}
(*@\hl{int radius = seekBar.getProgress();}@*)
(*@\hl{if (radius < 1) \{}@*)
    (*@\hl{radius = 1;}@*)
(*@\hl{\}}@*)
Bitmap bitmap = BitmapFactory.decodeResource(getResources(), R.drawable.image);
(*@\hl{imageView.setImageBitmap(blur.gaussianBlur(radius, bitmap));}@*)
\end{lstlisting}
\\
% \bottomrule
% \begin{lstlisting}
% float x = se.values[0];
% float y = se.values[1];
% float z = se.values[2];
% mAccelLast = mAccelCurrent;
% mAccelCurrent = (float) Math.sqrt((double) (x*x + y*y + z*z));
% float delta = mAccelCurrent - mAccelLast;
% mAccel = mAccel * 0.9f + delta; // perform low-cut filter
% \end{lstlisting}
% \vspace*{2em}
% \explanation{
%     \emph{Example C: Suggested Post-Processing.}
%     \begin{itemize}[leftmargin=*]
%         \item This code snippet (adapted from \cite{SO:shake}) calculates the device acceleration to detect a ``shake'' gesture.
%         \item The recommendation from a cluster of 2 methods suggests comparing the value with a predefined threshold and different procedures to start and stop the gesture handling.
%     \end{itemize}
% }
% &
% \begin{lstlisting}
% float x = sensorEvent.values[0];
% float y = sensorEvent.values[1];
% float z = sensorEvent.values[2];
% float mAccelLast = mAccelCurrent;
% mAccelCurrent = (float) Math.sqrt(x * x + y * y + z * z);
% float delta = mAccelCurrent - mAccelLast;
% mAccel = mAccel * 0.9f + delta;
% (*@\hl{if (mAccel > threshold) \{}@*)
%     (*@\hl{lastTimeShakeDetected = System.currentTimeMillis();}@*)
%     (*@\hl{isShaking = true;}@*)
%    (*@\hl{shakeListener.onShakeDetected();}@*)
% (*@\hl{\} else \{}@*)
%     (*@\hl{long timeDelta = (System.currentTimeMillis() - lastTimeShakeDetected);}@*)
%     (*@\hl{if (timeDelta > timeBeforeDeclaringShakeStopped \&\& isShaking) \{}@*)
%        (*@\hl{ isShaking = false;}@*)
%         (*@\hl{shakeListener.onShakeStopped();}@*)
%    (*@\hl{ \}}@*)
% (*@\hl{\}}@*)
% \end{lstlisting}
% \\

\bottomrule

\begin{lstlisting}
EditText et = (EditText)findViewById(R.id.inbox);
et.setSelection(et.getText().length());
\end{lstlisting}
\vspace*{0.5em}
\explanation{
    \emph{Example C: Correlated Statements.}
    \begin{itemize}[leftmargin=*]
        \item This code snippet moves the cursor to the end in a text area.\tnote{5} % \cite{SO:edittext}
        \item The recommended code suggests also configuring the action bar to create a more focused view.
        \item \textbf{Intersected from a cluster of 2 methods}.\tnote{6} % ~\cite{GH:intro11, GH:intro12}
    \end{itemize}
}
&
\begin{lstlisting}
(*@\hl{super.onCreate(savedInstanceState);}@*)
(*@\hl{setContentView(R.layout.material\_edittext\_activity\_main);}@*)
(*@\hl{getSupportActionBar().setDisplayHomeAsUpEnabled(true);}@*)
(*@\hl{getSupportActionBar().setDisplayShowTitleEnabled(false);}@*)
EditText singleLineEllipsisEt = (EditText) findViewById(R.id.singleLineEllipsisEt);
singleLineEllipsisEt.setSelection( singleLineEllipsisEt.getText().length());
\end{lstlisting}
% https://github.com/cymcsg/UltimateAndroid/blob/master/deprecated/UltimateAndroidGradle/demoofui/src/main/java/com/marshalchen/common/demoofui/sampleModules/MaterialEditTextActivity.java#L14
% https://github.com/klinker24/Android-SlidingMessaging/blob/master/src/main/java/com/klinker/android/messaging_sliding/scheduled/NewScheduledSms.java#L105
\\

\bottomrule

\begin{lstlisting}
PackageInfo pInfo = getPackageManager().getPackageInfo(getPackageName(), 0);
String version = pInfo.versionName;
\end{lstlisting}
% (*@\hl{try \{}@*)
%     PackageInfo pInfo = getPackageManager().getPackageInfo(getPackageName(), 0);
%     String version = pInfo.versionName;
%     (*@\hl{TextView versionView = (TextView) findViewById(R.id.about\_project\_version);}@*)
%     (*@\hl{versionView.setText("v" + version);}@*)
% (*@\hl{\} catch (PackageManager.NameNotFoundException ex) \{}@*)
%     (*@\hl{Log.e(TAG, getString(R.string.about\_error\_version\_not\_found));}@*)
% (*@\hl{\}}@*)
\vspace*{0.5em}
\explanation{
    \emph{Example D: Exact Recommendations.}
    \begin{itemize}[leftmargin=*]
        \item This partial code snippet gets the current version of the application. The rest of the code snippet (not shown) catches and handles possible \texttt{NameNotFound} errors.\tnote{7} % (truncated from \cite{SO:buildversion})
        \item The recommended code suggests the exact same error handling as in the original code snippet.
        \item \textbf{Intersected from a cluster of 2 methods}.\tnote{8} % ~\cite{GH:E1, GH:E2}
    \end{itemize}
}
&
\begin{lstlisting}
(*@\hl{try \{}@*)
    PackageInfo pInfo = getPackageManager().getPackageInfo(getPackageName(), 0);
    String version = pInfo.versionName;
    (*@\hl{TextView versionView = (TextView) findViewById(R.id.about\_project\_version);}@*)
    (*@\hl{versionView.setText("v" + version);}@*)
(*@\hl{\} catch (PackageManager.NameNotFoundException ex) \{}@*)
    (*@\hl{Log.e(...);}@*)
(*@\hl{\}}@*)
\end{lstlisting}
% https://github.com/front-line-tech/background-service-lib/blob/master/SampleService/servicelib/src/main/java/com/flt/servicelib/AbstractPermissionExtensionAppCompatActivity.java#L53
% https://github.com/brarcher/video-transcoder/blob/master/app/src/main/java/protect/videotranscoder/activity/MainActivity.java#L1597
\\

\bottomrule

\begin{lstlisting}
i.putExtra("parcelable_extra", (Parcelable) myParcelableObject);
\end{lstlisting}
% i.putExtra("parcelable_extra", (Parcelable) myParcelableObject);
% (*@\hl{i.putExtra("serializable\_extra", (Serializable) myParcelableObject);}@*)
\vspace*{0.5em}
\explanation{
    \emph{Example E: Alternative Recommendations.}
    \begin{itemize}[leftmargin=*]
        \item This partial code snippet demonstrates one way to attach an object to an \texttt{Intent}. The rest of the code snippet (not shown) shows a different way to serialize and attach an object.\tnote{9} % (adapted from \cite{SO:intent})
        \item \textbf{Intersected from a cluster of 10 methods}.\tnote{10}%~\cite{GH:F1,GH:F2,GH:F3,GH:F4}
    \end{itemize}
}
&
\begin{lstlisting}
(*@\hl{Intent intent = new Intent(this, BoardTopicActivity.class);}@*)
intent.putExtra(SMTHApplication.BOARD_OBJECT, (Parcelable) board);
(*@\hl{startActivity(intent);}@*)
\end{lstlisting}
\vspace*{1em}
\begin{itemize}[leftmargin=*]
    \item The recommended code does not suggest the other way of serializing the object, but rather suggests a common way to complete the operation by starting an activity with an \texttt{Intent} containing a serialized object.
\end{itemize}
\\

\bottomrule
\end{tabular}

\begin{tablenotes}

\item[1] Adapted from the Stack Overflow post ``Can I underline text in an android layout?'' [\url{https://stackoverflow.com/questions/2394939}], by Anthony Forloney [\url{https://stackoverflow.com/users/166712}].
\item[2] Adapted from \url{https://github.com/tonyvu2014/android-shoppingcart/blob/master/demo/src/main/java/com/android/tonyvu/sc/demo/ProductActivity.java}.
\item[3] Adapted from the Stack Overflow post ``How to set a bitmap from resource'' [\url{https://stackoverflow.com/questions/4955305}], by xandy [\url{https://stackoverflow.com/users/109112}].
\item[4] Adapted from \url{https://github.com/TonnyL/GaussianBlur/blob/master/app/src/main/java/io/github/marktony/gaussianblur/MainActivity.java}.
\item[5] Adapted from the Stack Overflow post ``Place cursor at the end of text in EditText'' [\url{https://stackoverflow.com/questions/6624186}], by Marqs [\url{https://stackoverflow.com/users/400493}].
\item[6] Adapted from \url{https://github.com/cymcsg/UltimateAndroid/blob/master/deprecated/UltimateAndroidGradle/demoofui/src/main/java/com/marshalchen/common/demoofui/sampleModules/MaterialEditTextActivity.java}.
\item[7] Adapted from the Stack Overflow post ``How to get the build/version number of your Android application?'' [\url{https://stackoverflow.com/questions/6593822}], by plus- [\url{https://stackoverflow.com/users/709635}].
\item[8] Adapted from \url{https://github.com/front-line-tech/background-service-lib/blob/master/SampleService/servicelib/src/main/java/com/flt/servicelib/AbstractPermissionExtensionAppCompatActivity.java}.
\item[9] Adapted from the Stack Overflow post ``How to send an object from one Android Activity to another using Intents?'' [\url{https://stackoverflow.com/questions/2141166}], by Jeremy Logan [\url{https://stackoverflow.com/users/76835}].
\item[10] Adapted from \url{https://github.com/zfdang/zSMTH-Android/blob/master/app/src/main/java/com/zfdang/zsmth_android/MainActivity.java}.

\item[] All Stack Overflow content is licensed under CC-BY-SA 3.0. All URLs are accessed in August 2018.

\end{tablenotes}

\end{threeparttable}
\end{table*}

% Reset listing style
\lstset{
	frame=tb,
    aboveskip=\medskipamount,
    belowskip=\medskipamount,
    basicstyle=\linespread{1}\scriptsize\ttfamily,
}

%% file: similar-code-study.tex
\section{The Opportunity for \aroma{}}
\label{sec:similar-code-study}

%\review{This entire section is new.}

\aroma{} is based on the idea that new code often resembles code that has already been written--- therefore, programmers can benefit from recommendations from existing code. To substantiate this claim, we conducted an experiment to measure the similarity of new code to existing code. This experiment was conducted on a large codebase in the Hack language.

We first collected all code commits submitted in a two-day period. From these commits, we extracted a set of changesets. A changeset is defined as a set of contiguous added or modified lines in a code commit. We filtered out changesets that were shorter than two lines or longer than seven lines. We decided to use this filter because longer changesets are more likely to span multiple methods, and we wanted to limit our dataset to code added or modified within a single method. Alternatively, we could have taken portions of changesets found within a single method; however, since changesets are raw text, finding the method boundaries involves additional parsing. We stuck to the simple solution of taking short changesets.
%\ksen{Could you mention what percentage of changesets were retained after this filtering?} \cb{I didn't keep track of this.}
%Alternatively, we could have taken longer changesets and extracted blocks of code (e.g. control statements). \ksen{Since you mentioned this, do you do this?  If not, why?} \fl{I don't think we have tried this. The criteria to split up changesets will be too arbitrary. I suggest we remove this sentence.}

For each of the first 1000 changesets in this set, we used \aroma{} to perform a code-to-code search, taking the snippet as input and returning a list of methods in the repository that contain structurally similar code.  \aroma{} was used because it was already implemented for Hack---but for the purpose of this experiment, any code-to-code search tool or clone detector can work. The results are ranked by similarity score: the percentage of features in the search query that are also found in the search result. For each changeset, we took the top-ranked method and its similarity score. 71 changesets did not yield any search result, because they contained only comments or variable lists, which \aroma{} disregards in search (see Section~\ref{sec:featurization}).
% 71 changesets contained no searchable features \ksen{Could you explain what "searchable features" mean?} and thus did not return any \aroma{} results---these were disregarded.
% For each changeset, we looked at the similarity score of the top result.

% \begin{figure}
% \centering
% \begin{minipage}{.5\textwidth}
%   \centering
%   \includegraphics[width=\linewidth]{sim_score_distribution}
%   \captionof{figure}{Distribution of Similarity Scores.}
%   \label{fig:sim_score_distribution}
% \end{minipage}%
% \begin{minipage}{.5\textwidth}
%   \centering
%   \includegraphics[width=.9\linewidth]{sim_score_plot}
%   \captionof{figure}{Plot of Changesets sorted by Similarity Scores.}
%   \label{fig:sim_score_plot}
% \end{minipage}
% \end{figure}

% We found that for 69.9\% of changesets, the most similar result had a score of at least 50\% as measured by \aroma{}, meaning the changeset shared at least half of its features with code already in another method. 17.5\% of changesets had a top \aroma{} result with a similarity score of at least 90\%. The distribution of similarity scores can be seen in ~\ref{fig:sim_score_distribution}, and a complete plot of the top similarity scores of all changesets can be seen in ~\ref{fig:sim_score_plot}. \ksen{I still do not know how to interpret 50\% similarity in features.  Does that imply that 50\% code could be copy-pasted?  How difficult would be to write the remaining 50\%?  I think a study of correleration between similarity score and programmer perception of similarity as done in Jose's paper would be convincing.}

\begin{figure}
\centering
\begin{minipage}{.45\textwidth}
  \centering
  \includegraphics[width=.85\linewidth]{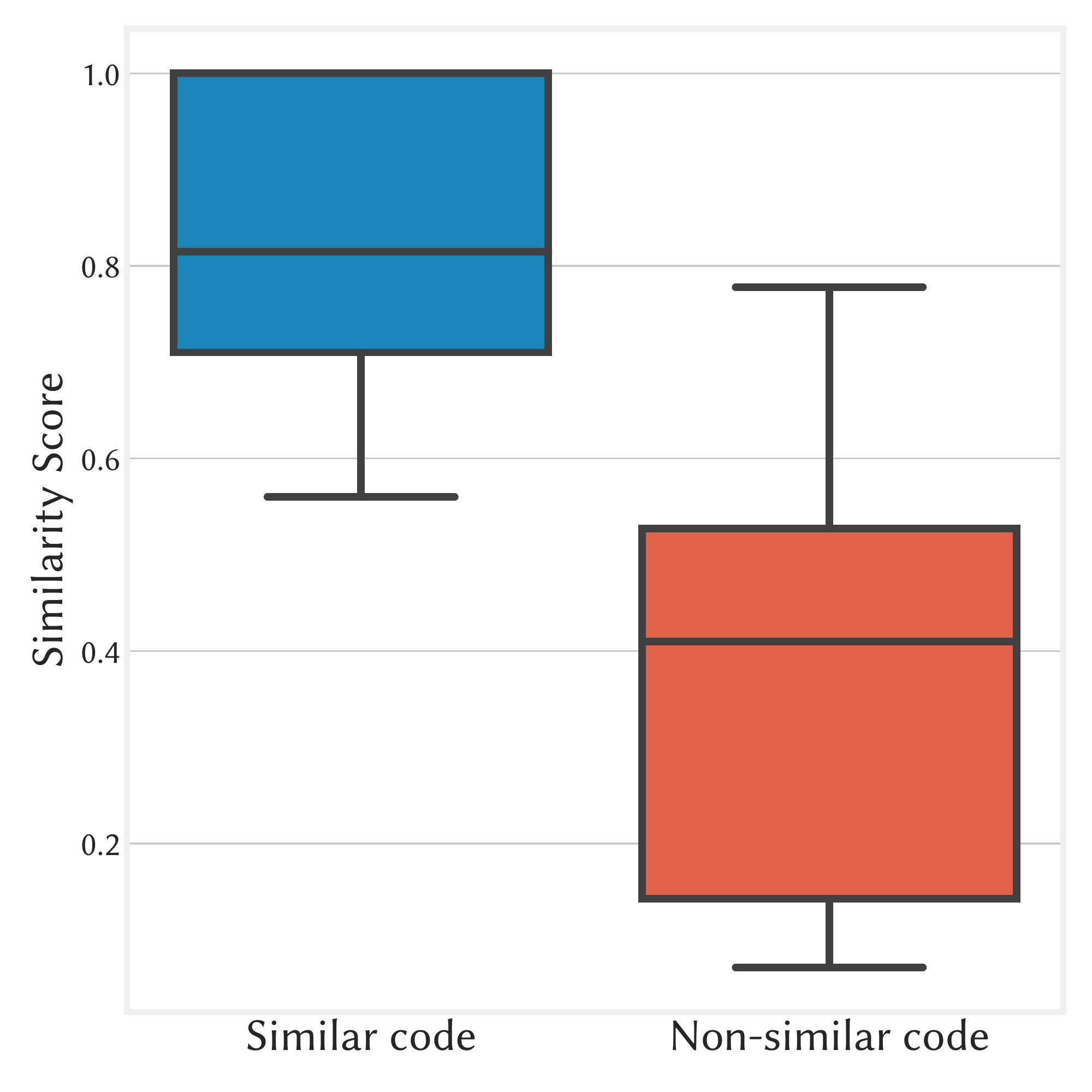}
  \captionof{figure}{Distribution of similarity scores used to obtain threshold}
  \label{fig:simboxplot}
\end{minipage} \hfill
\begin{minipage}{.5\textwidth}
  \centering
  \includegraphics[width=.7\linewidth]{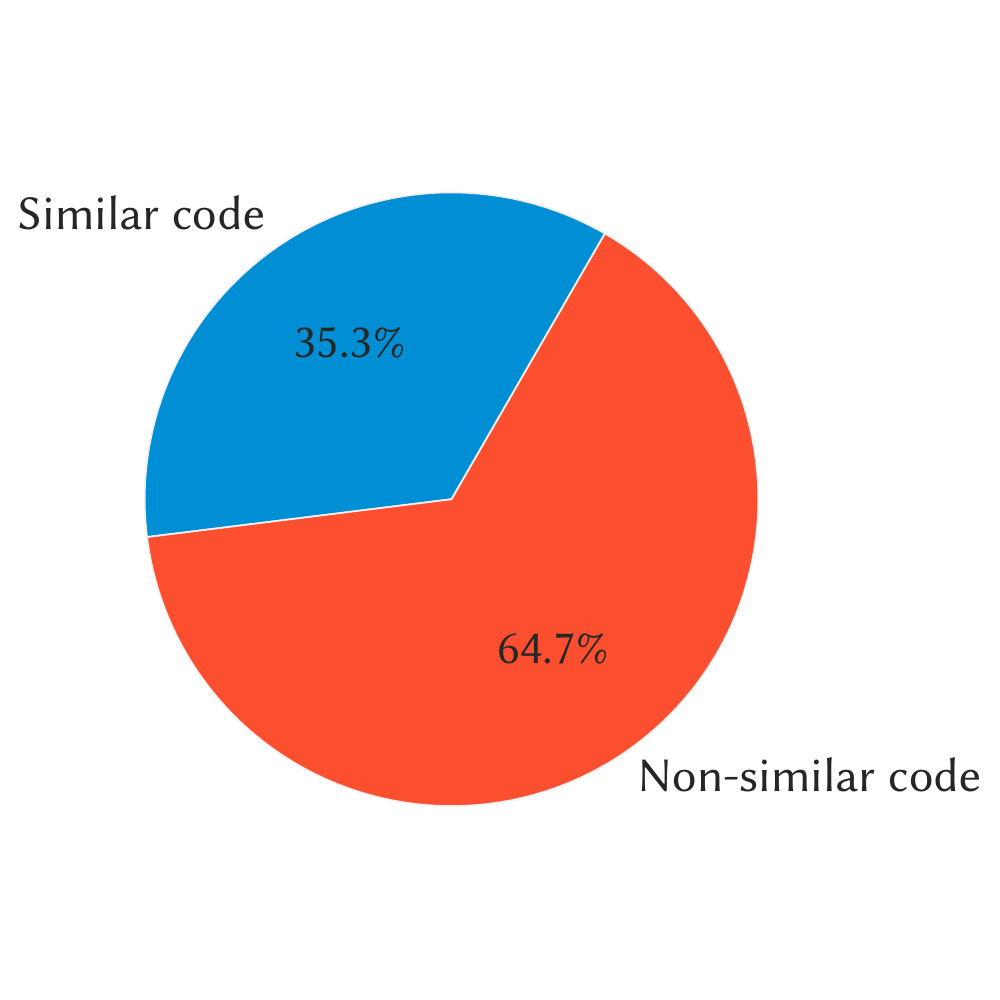}
  \captionof{figure}{Proportion of new code similar to existing code}
  \label{fig:sim_score_plot}
\end{minipage}
\end{figure}

To interpret the results, we first needed to assess the correlation between the similarity score (i.e. a measure of the syntactic similarity) and the semantic similarity between the changeset and the result. Two authors manually looked over a random sample of 50 pairs of changesets and result methods, and decided whether this method contained code similar enough to the changeset that a programmer could adopt the existing code (by copy-pasting or refactoring) with minimal changes. Using this criteria, each pair was deemed ``similar'' or ``not similar''. Conflicting judgments were cross-checked and re-assessed. As shown in the box plot in Figure~\ref{fig:simboxplot}, there is a clear distinction in similarity scores between the manually-labeled ``similar'' and ``not similar'' pairs. Note that in this figure, the top and bottom of the box represents the third and first quartile of the data. The lines extending above and below the box represent the maximum and minimum, and the line running through the box represents the median value.

We chose the first quartile of the similarity scores in the manually-labeled similar pairs---0.71---as the threshold similarity score to decide whether a retrieved code snippet contains meaningful semantic similarities to new code in the commit. We found that for 35.3\% of changesets, the most similar result had a score of at least 0.71, meaning that in these cases it would be easy for a programmer to adapt the existing code with minimal efforts, should the code be provided to them.

These results indicate that a considerable amount of new code contains similarities to code that already exists in a large code repository.
% These results indicate that new code frequently contains meaningful structural similarities to code that already exists in a large repository.
With \aroma{}, we aim to utilize this similar code to offer concise, helpful code recommendations to programmers. The amount of similar code in new commits suggests that \aroma{} has the potential to save a lot of programmers' time and effort.

%% file: algorithm.tex
\section{Algorithm} \label{sec:algorithm}

% child vs descendant

\begin{figure}[!ht]
\includegraphics[width=0.75\textwidth]{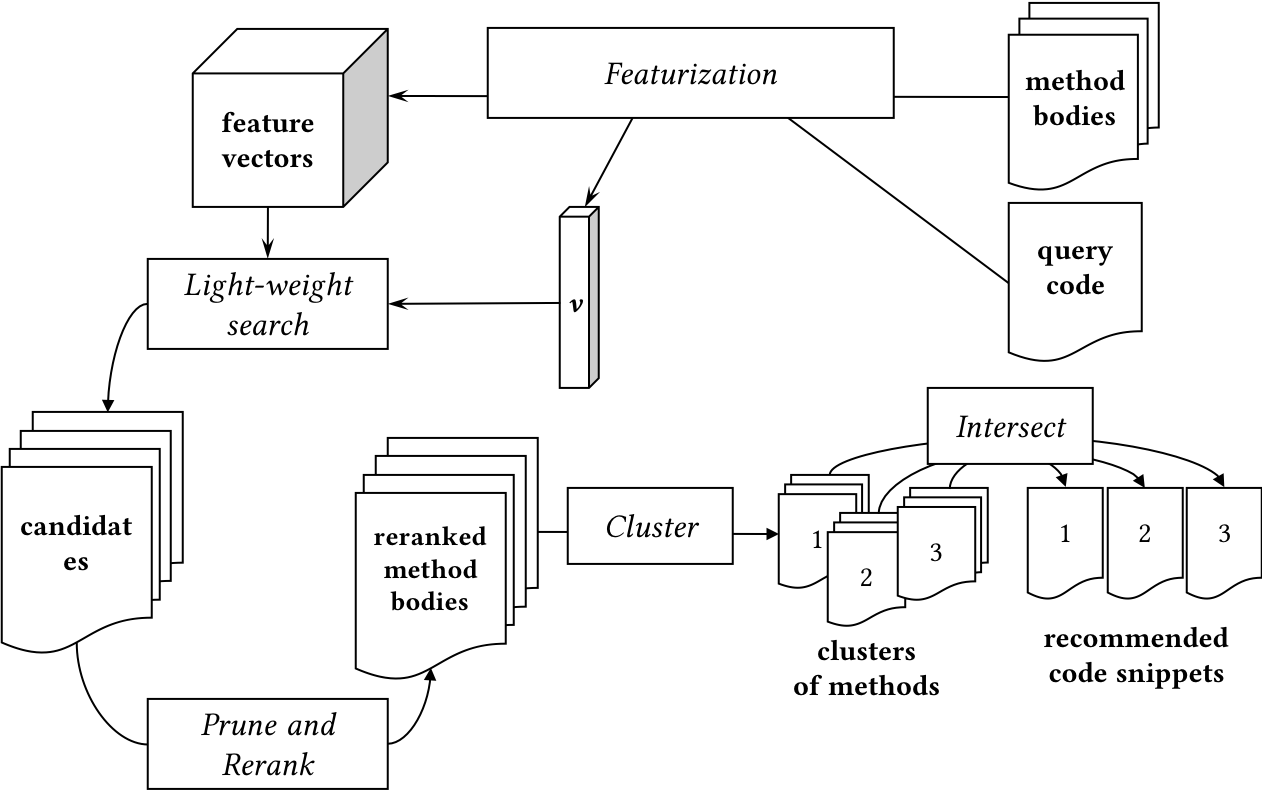}
\caption{\aroma{} code recommendation pipeline}%\fl{Updated figure.}}
% \ksen{The figure should be changed to refelect the changes I made to the beginning of section 3.}
\label{fig:aroma_model}
\end{figure}

Figure~\ref{fig:aroma_model} illustrates the overall architecture of \aroma{}.
%\fl{Listing, not Figure}
In order to generate code recommendations, \aroma{} must first featurize the code corpus. To do so, \aroma{} parses the body of each method in the corpus and creates its parse tree.  It extracts a set of structural features from each parse tree.
% %Such features mostly include bigrams and unigrams.  Unigrams are the non-keyword tokens present in the method body such as literals, method names, and property names.  Bigrams are formed by concatenating keyword or non-keyword tokens with non-keyword tokens.  For example, one could combine an integer literal $0$ with the label of its parent or grand-parent node in the parse tree, such as \texttt{+}, to get a bigram token. \fl{Instead of saying unigrams and bigrams, we probably just want to refer to the four kinds of features as we defined later. Parent features are technically not bigrams. Also, non-keyword tokens are not defined yet.}
% \item \textbf{Vectorization}. The features of a method body are represented as a sparse binary vector whose i$^{\rm th}$ entry is 1 if feature i is present in the method body and 0 otherwise.  The feature vectors of all method bodies are represented as a sparse matrix whose j$^{\rm th}$ row represents the feature vector of the j$^{\rm th}$ method body in the corpus.
% \end{itemize}
Then, given a query code snippet, \aroma{} runs the following phases to create recommendations:

\begin{itemize}
\item \textbf{Light-weight Search.} In this phase, \aroma{} takes a query code snippet, and outputs a list of the top few (e.g. 1000) methods that have the most overlap with the query.  To do so, \aroma{} extracts custom features from the query and each method in the corpus.  \aroma{} intersects the set of features of the query and each method body, and uses the cardinality of the intersection to compute the degree of overlap between the query and the method body.  To make this computation efficient, \aroma{} represents the set of features of a code snippet as a sparse vector and performs matrix multiplication to compute the degree of overlap of all methods with the query code.

%We next describe the formal details of the phase.

% It then selects a set of method bodies from the corpus such that query code's feature set has the most overlap with the feature sets of the selected method bodies.
% The selection is done by taking dot product of query code's feature vector with the feature vector of each method body in the corpus.  (The technique for creating these feature vectors, which are sparse, is described in Section~\ref{sec:reco-algo}.)
% % The selection is done by converting the set of features of each code snippet into a 0-1 sparse feature vector and by taking the dot product of query code's feature vector with the feature matrix.
% % The top $\eta_1$ method bodies whose dot products are highest are retrieved as the candidate set for recommendation.
% Even though the code corpus could contain millions of methods, this retrieval is fast due to efficient implementations of dot products of sparse vectors and matrices.
\item \textbf{Prune and Rerank.} In this phase, \aroma{} reranks the list of method bodies retrieved from the previous phase using a more precise, but expensive algorithm for computing similarity.

% Since the features of a code snippet are an over-approximate representation of that snippet, the ranking of the code snippets based on feature set overlap may not accurately rank the retrieved code snippets according to their closeness to the query code.  Moreover, the retrieved method bodies may contain statements that are not relevant to the query code.  Therefore, we need to rerank the retrieved code snippets so that method bodies that contain more parts of the query snippet are ranked higher than the method bodies that contain fewer parts of the query snippet. \todo{This explanation is not very clear.}
% In this step, \aroma{} ranks the retrieved code snippets from the previous phase based on their similarity to the query snippet.
% For the ranking, \aroma{} prunes each retrieved code snippet so that the resulting snippet becomes maximally similar
% %\fl{becomes more similar?}
% to the query snippet.  The similarity metric used for this pruning is based on a modified formulation of the Jaccard distance between the features of the query and that of the pruned code snippet.
\item \textbf{Cluster and Intersect.} In the final phase, \aroma{} clusters the reranked list of code snippets from the previous phase.  Clustering is based on the similarity of the method bodies.  Clustering also needs to satisfy constraints which ensure that recommendations are of high quality. Therefore, we have devised a custom clustering algorithm which takes the constraints into account.  \aroma{} then intersects the snippets in each cluster to come up with recommended code snippets.  This approach of clustering and intersection helps to create a succinct, yet diverse set of recommendations.
\end{itemize}

We next describe the details of each step using the code snippet shown in Listing~\ref{lst:example_code} as the running example.

% \cite{SO:iterate_view}
\begin{lstlisting}[label={lst:example_code}, caption={A code snippet adapted from a Stack Overflow post.\protect\footnotemark} This snippet is used as the running example through Section~\ref{sec:algorithm}.]
if (view instanceof ViewGroup) {
    for (int i = 0; i < ((ViewGroup) view).getChildCount(); i++) {
        View innerView = ((ViewGroup) view).getChildAt(i);
    }
}
\end{lstlisting}
\footnotetext{Adapted from the Stack Overflow post ``How to hide soft keyboard on android after clicking outside EditText?'' [\url{https://stackoverflow.com/questions/11656129}], by Navneeth G [\url{https://stackoverflow.com/users/1135909}]. CC-BY-SA 3.0 License. Accessed in August 2018.}

\subsection{Definitions}

In this section, we introduce several notations and definitions used to compute the features of a code snippet.  The terminologies and notations are also used to describe \aroma{} formally.

\begin{definition}[Keyword tokens] This is the set of all tokens in a language whose values are fixed as part of the language. Keyword tokens include keywords such as \texttt{while}, \texttt{if}, \texttt{else}, and symbols such as \texttt{\{}, \texttt{\}}, \texttt{.}, \texttt{+}, \texttt{*}. The set of all keyword tokens is finite for a language.
\end{definition}

\begin{definition}[Non-keyword tokens] This is the set of all tokens that are not keyword tokens.  Non-keyword tokens include variable names, method names, field names, and literals.  \end{definition}

\noindent Examples of non-keyword tokens are \texttt{i}, \texttt{length}, $0$, $1$, etc.  The set of non-keyword tokens is non-finite for most languages.

\begin{definition}[Simplified Parse Tree]  A simplified parse tree is a data structure we use to represent a program.  It is recursively defined as a non-empty list whose elements could be any of the following:
    \begin{itemize}
        \item a non-keyword token,
        \item a keyword token, or
        \item a simplified parse tree.
    \end{itemize}
    Moreover, a simplified parse tree cannot be a list containing a single simplified parse tree.
\end{definition}

We picked this particular representation of programs instead of a conventional abstract syntax tree representation because the representation only consists of program tokens, and does not use any special language-specific rule names such as \texttt{IfStatement}, \texttt{block} etc.  As such, the representation can be used uniformly across various programming languages.  Moreover, one could perform an in-order traversal of a simplified parse tree and print the token names to obtain the original program, albeit unformatted.  We use this feature of a simplified parse tree to show the recommended code snippets.

\begin{definition}[Label of a Simplified Parse Tree]  The label of a simplified parse tree is obtained by concatenating all the elements of the list representing the tree as follows:
    \begin{itemize}
        \item If an element is a keyword token, the value of the token is used for concatenation.
        \item If an element is a non-keyword token or a simplified parse tree, the special symbol \texttt{\#} is used for concatenation.
    \end{itemize}
\end{definition}

\noindent For example, the label of the simplified parse tree \texttt{["x", ">", ["y", ".", "f"]]} is \texttt{"\#>\#"}.

\begin{figure}[!ht]
\includegraphics[width=0.75\textwidth]{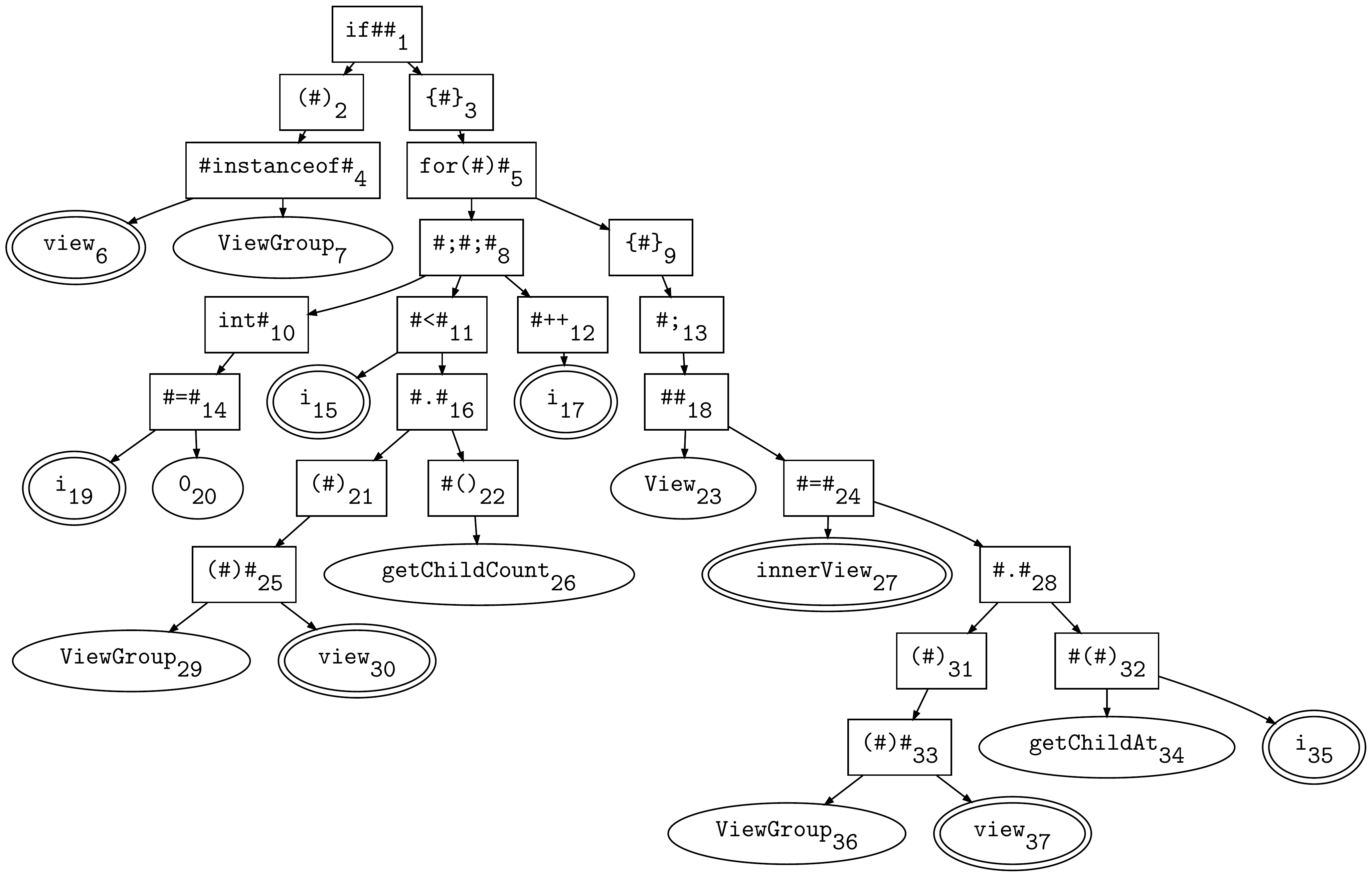}
\caption{The simplified parse tree representation of the code in Listing~\ref{lst:example_code}. Keyword tokens at the leaves are omitted to avoid clutter. Variable nodes are highlighted in double circles.}
\label{fig:example_tree}
\end{figure}

Figure~\ref{fig:example_tree} visualizes the simplified parse tree of the code snippet in Listing~\ref{lst:example_code}. In the figure, each internal node represents a simplified parse tree, and is labeled using the tree's label as defined above.  Since keyword tokens in a simplified parse tree become part of the label of the tree, we do not create leaf nodes for keyword tokens in the tree diagram---we only add leaf nodes for non-keyword tokens.  We show the label of each node in the tree, and add a unique index to each label as subscript to distinguish between any two similar labels.

To obtain the simplified parse tree of a code snippet, \aroma{} relies on a language-specific parser. For example, \aroma{} utilizes the \texttt{ANTLR4}~\cite{ANTLR} Java parser to produce the parse tree for a Java program.  \aroma{} traverses the parse tree produced by the parser to collect at each internal node the tokens and subtrees that are immediate descendants of each internal node of the parse tree.  The collected elements at each node form a list, which is a simplified parse tree.  \aroma{} uses the list at each internal node to create the label for the node.  The distinction between keyword and non-keyword tokens is done using the language's lexical specification.  \aroma{} performs a second traversal of the tree and uses the static scoping rules of the language to identify the global variables.  \aroma{} uses the knowledge of the global variables in the featurization phase which we describe later.  Note that this process of creating a simplified parse tree from a code snippet is language-dependent and requires knowledge about the grammar and static scoping rules of the language.  Once the simplified parse tree of a code snippet has been created, the rest of \aroma{}'s algorithm is language-agnostic.

%\cb{$\texttt{view}_{30}$ is technically the fourth child of $\texttt{(\#)\#}$---however, since we omit the leaf keyword tokens, it makes more sense to say it is the second child (Michael commented on this as well)} \ksen{It might be difficult to change to 2 from 4 since we have to make a several changes throughout the paper to be consistent.}

Given a simplified parse tree $t$, we use the following notations. All examples refer to Figure~\ref{fig:example_tree}.

% \ksen{Give example for each of the following notations using Figure 3}
\begin{itemize}
    \item $L(t)$ denotes the label of the tree $t$. E.g. $L(\texttt{if\#\#}_1) = \texttt{if\#\#}$.
    \item $N(t)$ denotes the list of all non-keyword tokens present in $t$ or in any of its sub-trees, in the same order as appearing in the source code. E.g. $N(\texttt{\#.\#}_{28}) = [\texttt{ViewGroup}_{36}, \texttt{view}_{37}, \texttt{getChildAt}_{34}, \texttt{i}_{35}]$.
    %\fl{Changed to list because we rely on this ordering when defining $C(n)$ for method invocations.}
    \item If $n$ is a non-keyword direct child of $t$, then we use $P(n)$ to denote the parent of $n$ which is $t$. E.g. $P(\texttt{view}_6) = \texttt{\#instanceof\#}_4$.
    \item If $t'$ is a simplified parse tree and is a direct child of $t$, then we again use $P(t')$ to denote the parent of $t'$ which is $t$. E.g. $P(\texttt{\#instanceof\#}_4) = \texttt{(\#)}_2$.
    \item If $n_1$ and $n_2$ are two non-keyword tokens in a program and if $n_2$ appears after $n_1$ in the program without any intervening non-keyword token, then we use ${\tt Prev}(n_2)$ to denote $n_1$ and ${\tt Next}(n_1)$ to denote $n_2$. E.g. ${\tt Prev}(\texttt{view}_{30}) = \texttt{ViewGroup}_{29}, {\tt Next}(\texttt{ViewGroup}_{29}) = \texttt{view}_{30}$.
    \item If $n_1$ and $n_2$ are two non-keyword tokens denoting the same local variable in a program and if $n_1$ and $n_2$ are the two consecutive usages of the variable in the source code, then we use ${\tt PrevUse}(n_2)$ to denote $n_1$ and ${\tt NextUse}(n_1)$ to denote $n_2$. E.g. ${\tt PrevUse}(\texttt{view}_{30}) = \texttt{view}_6, {\tt NextUse}(\texttt{view}_{30}) = \texttt{view}_{37}$.
    \item If $n$ is a non-keyword token denoting a local variable and it is the $i^{\rm th}$ child of its parent $t$, then the context of $n$, denoted by $C(n)$, is defined to be:
    \begin{itemize}
        \item $(i, L(t))$, if $L(t) \neq \texttt{\#.\#}$. E.g. $C(\texttt{view}_{30}) = (2, \texttt{(\#)\#})$.
        \item The first non-keyword token that is not a local variable in $N(t)$, otherwise. This is to accommodate for cases like \texttt{x.foo()}, where we want the context feature for \texttt{x} to be \texttt{foo} rather than $(1, \texttt{\#.\#})$, because the former better reflects its usage context.
        %\fl{Revised this definition and added why we use the first non-variable leaf token.}
    \end{itemize}
% \begin{equation}
% C(n ) = \left\{\begin{matrix}
% (i, L(P(n))), \mbox{ if }L(P(n)) \neq \texttt{\#.\#} \\
% \mbox{The first non-keyword token that is not a local variable in } N(P(n)), \mbox{ otherwise}
% \end{matrix}\right.
% \end{equation}
\end{itemize}

\subsection{Featurization}
\label{sec:featurization}

The high-level goal of this step is to take a simplified parse tree for a code snippet, and extract a set of structural features from that parse tree. A key requirement of the features is that if two code snippets are similar, they should have the same collection of features.

A simple way to featurize a code snippet is to treat the labels of all nodes in the simplified parse tree as features.  This simple approach creates problem if we have two code snippets 1) which only differ in local variable names, and 2) where one code snippet can be obtained from the other by alpha renaming the local variables.  The two code snippets should be considered as similar, but the collection of features will differ in the name of some of the variables.  Therefore, we replace each token that denotes a local variable by a special token \texttt{\#VAR}.  We do not perform similar replacements for global variables and method names.  This is because such identifiers are often part of some library API and cannot be alpha-renamed to obtain similar programs.

Treating the labels of parse tree nodes as the only features does not help to capture the relation between the nodes.  Such relations are necessary to identify the structural features of a code snippet.  For example, without such a relation \aroma{} will treat the snippets \texttt{if (x > 0)  z = 3;} and \texttt{if (z > 3)  x = 0;} as similar since they have the exact same collection of node labels (i.e. \texttt{\{if\#\#, (\#), \#>\#, \#;, \#=\#, 0, 3, \#VAR, \#VAR\}}).  If we can somehow create a feature encapsulating the fact that \texttt{3} belongs to the body of the first \texttt{if} statement, \aroma{} will distinguish between the two snippets.  Therefore, \aroma{} also creates features which represent some relations between certain pairs of nodes in the parse tree.  Examples of some such features involving the token $3$ are \texttt{(if\#\#,2,3)}, \texttt{(\#=\#,2,3)}, and \texttt{(\#VAR,3)}.  The first feature, which is denoted as a triplet, states that the $2^{\rm nd}$ child of a node labeled \texttt{if\#\#} has a descendant leaf node with label \texttt{3}.  Similarly, the second feature asserts that the $2^{\rm nd}$ child of a node labeled \texttt{\#=\#} has a descendant leaf node with label \texttt{3}.  We call these two features \emph{parent features}, as they help capture the relation of a leaf node with its parent, grand-parent, and great-grand parent.  The third feature relays the fact that a variable leaf node appears before $3$.  We call such features \emph{sibling features}.
In summary, the parent features and sibling features capture some local relations between the nodes in a parse tree.  However, these features are not exhaustive enough to recreate the parse tree from the features.  These non-exhaustiveness of features helps \aroma{} tolerate some non-similarities in otherwise similar code snippets, and helps \aroma{} to retrieve some closely related, but nonidentical code snippets during search.

Since we replace all local variable names with \texttt{\#VAR}, we also need to relate two variable usages in a code snippet which refer to the same local variable.  For example, in the code snippet \texttt{if (y < 0) x = -x;}, we will have three features of the form \texttt{\#VAR} corresponding to the two occurrences of \texttt{x} and one occurrence of \texttt{y}.  However, the collection of features described so far does not express the fact that two \texttt{\#VAR} features refer to the same variable.  There is no direct way to state two variables are related since we have gotten rid of variable names.  Rather, we capture features about the fact that the \emph{consecutive usage context} of the same local variables are related.  We call such features \emph{variable usage features.}  For example, the context of the two usages of \texttt{x} are \texttt{(1,\#=\#)} and \texttt{(1,-\#)}, respectively.  The first context corresponds to the first usage of \texttt{x} and denotes that there is a variable which is the first child of the node labeled \texttt{\#=\#}.  The index and the parent node label together forms the context of this particular variable usage.  Similarly, the second context denotes the second usage of \texttt{x}.  We create a feature of the form \texttt{((1,\#=\#),(1,-\#))} which captures the relation between the context of two consecutive usage of the same variable.

We now describe formally how a code snippet is featurized by \aroma{}.  Given a simplified parse tree, we extract four kinds of features for each non-keyword token $n$ in the program represented by the tree:

\begin{enumerate}[topsep=0.5\baselineskip]
    \item A \emph{Token Feature} of the form $n$.  If $n$ is a local variable, we replace $n$ with \texttt{\#VAR}.
    % \item \emph{Parent Features} of the form $(n, i_1, L(P(n)))$, $(n, i_2, L(P(P(n))))$, and $(n, i_3, L(P(P(P(n)))))$.  Here $P(P(n))$ is the $i_3^{\rm th}$ children of $P(P(P(n)))$, $P(n)$ is the $i_2^{\rm th}$ children of $P(P(n))$, and $n$ is the $i_1^{\rm th}$ children of $P(n)$.
    \item \emph{Parent Features} of the form $(n, i_1, L(t_1))$, $(n, i_2, L(t_2))$, and $(n, i_3, L(t_3))$.  Here $n$ is the $i_1^{\rm th}$ child of $t_1$, $t_1$ is the $i_2^{\rm th}$ child of $t_2$, and $t_2$ is the $i_3^{\rm th}$ child of $t_3$.  As before, if $n$ is a local variable, then we replace $n$ with \texttt{\#VAR}.  Note that in each of these features, we do not specify if the third element in a feature is the parent, grand-parent, or the great-grand parent.  This helps \aroma{} to tolerate some non-similarities in otherwise similar code snippets.
    % \fl{Replaced the nested $P(P(P(n)))$ with $t_1, t_2, t_3$. Should we say if there is no great-grand parent then there's no parent feature, or is that obvious?}
    \item \emph{Sibling Features} of the form $(n, {\tt Next}(n))$ and $({\tt Prev}(n), n)$. As before, if any of $n, {\tt Next}(n), {\tt Prev}(n)$ is a local variable, it is replaced with \texttt{\#VAR}.
    %\fl{Added special case handling for \texttt{\#VAR}.}
    \item \emph{Variable Usage Features} of the form $(C({\tt PrevUse}(n)), C(n))$ and $(C(n), C({\tt NextUse}(n)))$.  We only add these features if $n$ is a local variable.
\end{enumerate}

For a non-keyword token $n \in N(t)$, we use $F(n)$ to denote the multi-set of features extracted for $n$.  We extend the definition of $F$ to a set of non-keyword tokens $Q$ as follows: $F(Q) = \uplus_{n \in Q} F(n)$ where $\uplus$ denotes multi-set union.  For a simplified parse tree $t$, we use $F(t)$ to denote the multi-set of features of all non-keyword tokens in $t$, i.e. $F(t) = F(N(t))$.  Let $\mathcal{F}$ be the set of all features that can extracted from a given corpus of code.
%\fl{Added definition for a set of nodes, $F(Q)$.}

Table~\ref{tab:query_example_features} illustrates the features extracted for two non-keyword tokens from the simplified parse tree in Figure~\ref{fig:example_tree}.  In the interest of space, we do not show the features extracted by \aroma{} for all non-keyword tokens.

\begin{table}[!ht]
\caption{Features for selected tokens in Figure~\ref{fig:example_tree}}
\centering
\scriptsize
\setlength{\tabcolsep}{0pt}
\begin{tabular}{@{}m{0.1\textwidth}m{0.15\textwidth}m{0.15\textwidth}m{0.2\textwidth}m{0.2\textwidth}@{}}
% \begin{tabular}{@{}lllll@{}}
\toprule
& Token Feature & Parent Features & Sibling Features & Variable Usage Features \\
% & (1) & (2) & (3) & (4) \\
\midrule
% view & \#VAR &
%   \begin{tabular}{@{}l@{}}
%         (\#VAR, 1, \#instanceof\#) \\
%         (\#VAR, 1, \#) \\
%         (\#VAR, 1, if\#\#)
%     \end{tabular}
%     & (\#VAR, ViewGroup) & ((1, \#instanceof\#), (2, (\#)\#)) \\
$\texttt{view}_{30}$ & \texttt{\#VAR} &
    \begin{tabular}{@{}l@{}}
        (\texttt{\#VAR}, 2, \texttt{(\#)\#}) \\
        (\texttt{\#VAR}, 1, \texttt{(\#)}) \\
        (\texttt{\#VAR}, 1, \texttt{\#.\#})
    \end{tabular}
    &
    \begin{tabular}{@{}l@{}}
        (\texttt{ViewGroup}, \texttt{\#VAR}) \\
        \parbox{0.15\textwidth}{
            (\texttt{\#VAR}, \\ \hspace*{0.5em} \texttt{getChildCount})
        }
    \end{tabular}
    &
    \begin{tabular}{@{}l@{}}
        \parbox{0.25\textwidth}{
            ((1, \texttt{\#instanceof\#}), \\ \hspace*{0.5em} (2, \texttt{(\#)\#}))
        } \\
        ((2, \texttt{(\#)\#}), (2, \texttt{(\#)\#}))
    \end{tabular}
    \\
% \midrule
% i & \#VAR &
%   \begin{tabular}{@{}l@{}}(\#VAR, 1, \#<\#)\\ (\#VAR, 2, \#;\#;\#) \\
%                       (\#VAR, 1, for(\#)\#)\\ (\#VAR, 1, \{\#\}) \\
%                       (\#VAR, 2, if\#\#)\end{tabular}
%   & (\#VAR, \#\.\#)
%   & \begin{tabular}{@{}l@{}}((1, \#=\#), (1, \#<\#)) \\
%               ((1, \#<\#), (1, \#++))\end{tabular}\\
\midrule
$0_{20}$ & 0 &
    \begin{tabular}{@{}l@{}}
        (0, 2, \texttt{\#=\#}) \\
        (0, 1, \texttt{int\#}) \\
        (0, 1, \texttt{\#;\#;\#;})
    \end{tabular}
    &
    \begin{tabular}{@{}l@{}}
    (\texttt{\#VAR}, 0) \\
    (0, \texttt{\#VAR})
    \end{tabular}
    & - \\
\bottomrule
\end{tabular}
\label{tab:query_example_features}
\end{table}

\subsection{Recommendation Algorithm}
\label{sec:reco-algo}

\subsubsection{Phase I:  Light-weight Search}

In this phase, \aroma{} takes a query code snippet, and outputs a list of the top few (e.g. 1000) methods that contain the most overlap with the query.  To compute the top methods, we need to compute the degree of overlap between the query and each method body in the corpus. Because our corpus has millions of methods, we need to make sure that the degree of overlap can be computed fast at query time. We use the feature sets of code snippets to compute the degree of overlap, which we call the \emph{\bf overlap score}.  Specifically, \aroma{} intersects the set of features of the query and the method body, and uses the cardinality of the intersection as the overlap score.  Computing intersection and its cardinality could be computationally expensive.  For efficient computation, we represent the set of features of a code snippet as a sparse vector and perform matrix multiplication to compute the overlap score of all methods with the query code.  We next describe the formal details of the phase.

Given a large code corpus containing millions of methods.  \aroma{} parses and creates a simplified parse tree for each method body.  It then featurizes each simplified parse tree. Let $M$ be the set of simplified parse trees of all method bodies in the corpus. \aroma{} also parses the query code snippet to create its simplified parse tree, say $q$, and extracts its features.  For the simplified parse tree $m$ of each method body in the corpus, we use the cardinality of the set $S(F(m)) \cap S(F(q))$ as an approximate score, called \emph{overlap score}, of how much of the query code snippet overlaps with the method body.  Here $S(X)$ denotes the set of elements of the multi-set $X$, where we ignore the count of each element in the multi-set. \aroma{} computes a list of $\eta_1$ method bodies whose overlap scores are highest with respect to the query code snippet.  In our implementation $\eta_1$ is usually 1000. %\fl{Removed phase 1 score threshold $\tau_1$ because it is not used in our implementation.}

The computation of this list can be reduced to a simple multiplication between a matrix and a sparse vector as follows. The features of a code snippet can be represented as a sparse vector of length $|\mathcal{F}|$---the vector has an entry for each feature in $\mathcal{F}$.  If a feature $f_i$ is present in $F(m)$, the multi-set of features of the simplified parse tree $m$, then the $i^{\rm th}$ entry of the vector is 1 and 0 otherwise.  Note that the elements of each vector can be either 0 or 1---we ignore the count of each feature in the vector.  To understand this decision, consider a method $m$ that contains a feature $f$ numerous times (say $n$). Then, say we give \aroma{} a query $q$ that contains $f$ once. The overlap score between $m$ and $q$ will be increased by $n$, even though the multiple instances of this feature do not actually indicate greater overlap between $m$ and $q$.  The sparse feature vectors of all method bodies can then be organized as a matrix, say $D$, of shape $|M| \times |\mathcal{F}|$.  Let $v_q$ be the sparse feature vector of the query code snippet $q$.  Then $D \cdot v_q$ is a vector of size $|M|$ that gives the overlap score of each method body with respect to the query snippet.  \aroma{} picks the top $\eta_1$ method bodies with the highest overlap scores.  Let $N_1$ be the set of simplified parse trees of the method bodies picked by \aroma{}.

The corpus we used for evaluation has over 37 million unique features. But each method has an average of 63 methods, so the feature vectors are very sparse. Thus, the matrix multiplication described above can be done efficiently using a fast sparse matrix multiplication library---for our corpus, this phase finishes in less than a second.

%\fl{I want to extend the discussion of design choices here, regarding why we use 0-1 vs. count, and why we need pruning and reranking after this step.}
%The overlap score is designed to accommodate millions of methods, and enables \aroma{} to isolate a small set of method bodies that have the most overlap with the query.  However, the overlap score only approximation of the similarity score we will calculate in the next phase. Put simply, this step is fast but imprecise. It allows us to narrow the millions of methods in our corpus down to the thousand that contain the most overlap, and are thus the most likely to produce good recommendations. In the next step, we perform a more rigorous and labor-intensive reranking on the relatively short list of methods attained in this step.

%\ksen{An example showing how a list of method bodies needs to be reranked (as Michael suggested) would be useful.  Frank, can you try to write such an example?}
\subsubsection{Phase II: Prune and Rerank} \label{sec:pruning}

In the following phases, we need a sub-algorithm to compute a maximal code snippet that is common to two given code snippets.  For example, given the code snippets \texttt{x = 1; y = 2;} and \texttt{y = 2; z = 3;}, we need an algorithm that computes \texttt{y = 2;} as the intersection of the two code snippets.  This algorithm could be implemented using a longest-common subsequence (LCS)~\cite{Porter:1997} computation algorithm on strings by treating the two code snippets as strings.  Such an algorithm was used in SNIFF~\cite{Chatterjee:2009} (which performs natural language small code snippet search).  However, LCS does not work well for \aroma{} because often the common parts between two code snippets may not be exactly similar.  To illustrate this point, suppose we are given the two code snippets \texttt{x = 1; if (y > 1) if (z < 0) w = 4;} and \texttt{if (z < 0) if (y > 1) w = 4; v = 10;}, where we have swapped the nesting of the two \texttt{if} statements.  LCS will retrieve either \texttt{ if (y > 1)  w = 4;} or \texttt{if (z < 0)  w = 4; } as the intersection, i.e. LCS drops one of the \texttt{if} statements along with the non-common assignment statements.  Ideally, we should have both \texttt{if} statements in the intersection, i.e. the intersection algorithm should compute either \texttt{if (y > 1) if (z < 0) w = 4;} or \texttt{if (z < 0) if (y > 1) w = 4;} as the intersection.

The example also shows that we can have at most two intersected code snippets when fuzzy similarity exists between the given code snippets---a snippet will either be most similar to the first snippet, or the second snippet. We resolve this ambiguity by picking the intersected snippet that is close to the second snippet. Thus, we can think of the intersection as a code snippet obtained by taking the second snippet, and dropping its fragments which have no resemblance to the first snippet. That is, the algorithm is \emph{pruning} the second snippet while retaining the parts common with the first one.

A simple way to \emph{prune} the second snippet is to look at its parse tree and find a subtree which is most similar to the first snippet.  However, such an algorithm will be expensive because there are exponentially many subtrees in a given tree.  Instead, \aroma{} uses a greedy algorithm which gives us a maximal subtree of the second snippet's parse tree.  We have also observed that if we can identify all the leaf nodes in the second snippet's parse tree that need to be present in the intersection, we can get a maximal subtree by simply retaining all the nodes and edges in the tree that lie in a path from the root to the identified leaf nodes.  We next formally describe the pruning algorithm.

Let us assume we are given two code snippets, say $m_1$ and $m_2$, in the form of their parse trees.
The computation of the optimal pruned simplified parse tree, say $m_p$, requires us to find a subset, say $R$, of the leaf nodes of $m_2$.  Recall that the set of leaf nodes of $m$ is denoted by $N(m)$ and contains exactly the non-keyword tokens in the parse tree.    The set $R$ should be such that the similarity between $m_p$ and $m_1$ is maximal.  We will use the cardinality of the multi-set intersection of the features of two code snippets as their similarity score.  That is, the similarity score between two snippets given as parse trees, say $m_1$ and $m_2$, is $|F(m_1) \cap F(m_2)|$.  Let us denote it by ${\rm\it SimScore}(m_1, m_2)$.  Once we have computed the set of leaf nodes (i.e. $R$) that need to be present in the intersection, $m_p$ is the subtree consisting of the nodes in $R$, and any internal nodes and edges in $m_2$ which are along a path from any $n \in R$ to the root node in $m_2$.  The greedy algorithm for computing $R$ is described in Algorithm~\ref{alg:prune}.

\begin{algorithm}
\caption{The Pruning Algorithm}
\label{alg:prune}
\begin{algorithmic}[1]
\Procedure{Prune}{$F(m_1), m_2$}
\State \Comment The first argument takes the features of $m_1$ instead of $m_1$ itself to simplify the description of Phase III.
\State $R \gets \emptyset$
\State $F \gets \emptyset$
\Repeat
    \State $n \gets {\rm argmax}_{n' \in N(m_2) \setminus R} {\rm\it SimScore} (F(m_1), (F \uplus F(n')))$
    \If{$n$ exists and ${\rm\it SimScore}(F(m_1), F \uplus F(n)) > {\rm\it SimScore}(F(m_1), F)$}
        \State $R \gets R \cup \{n\}$
        \State $F \gets F \uplus F(n)$
    \EndIf
\Until{$R$ does not change anymore}
%\State $m_p \gets$ a subset of $m$, obtained by retaining all the non-keyword tokens in $R$ and any internal node or edge which appears in a path from a $n \in R$ to the root of $m_2$
\State \textbf{return}  $m_p$ where $m_p$ is obtained from $m_2$ by retaining all the non-keyword tokens in $R$ and
\State any internal node or edge which appears in a path from a $n \in R$ to the root of $m_2$
\EndProcedure
\end{algorithmic}
\end{algorithm}

% \vspace*{3ex}
% \noindent${\tt\bf Algorithm\  Prune}(F(m_1), m_2)$:
% \begin{enumerate}
%     \item $R \leftarrow \emptyset$. %\fl{Use another letter than $P$.}
%     \item $F \leftarrow \emptyset$.
%     \item Find $n$ such that \[ n = {\rm argmax}_{n' \in N(m_2)-R} {\rm\it SimScore} (F(m_1), (F \uplus F(n'))) \] and \[{\rm\it SimScore}(F(m_1), F \uplus F(n)) > {\rm\it SimScore}(F(m_1), F).\]
%     \item If such an $n$ exists, then $R \leftarrow R \cup \{n\}$ and $F \leftarrow F \uplus F(n)$.  Go back to Step 3.
%     \item Else return $m_p$ where $m_p$ is obtained from $m$ by retaining all the non-keyword tokens in $R$ and any internal node or edge which appears in a path from a $n \in R$ to the root of $m_2$.  Then ${\tt Prune}(F(m_1), m_2)$ is $m_p$.
%     \end{enumerate}

In the algorithm, \aroma{} maintains the collection of the features of the intersected snippet in the variable $F$.  The variable $R$ maintains the set of leaf nodes in the intersected code.  Initially, the algorithm starts with an empty set of leaf nodes. It then iteratively adds more leaf nodes to the set from the parse tree of the second method (i.e. $m_2$).  A node $n$ is added if it increases the similarity between the first method and the tree that can be obtained from $R$.  Since $F$ maintains the features of the tree that can be constructed from $R$, we can get the features of $R \cup \{n\}$ by simply adding the features associated with $n$ (i.e. $F(n)$) to $F$.  If such a node cannot be found, the algorithm constructs the intersected tree from $R$ and returns it.

We are next going to show how \aroma{} uses the pruning algorithm to rerank the snippets retrieved in Phase 1.  Given a query, say $q$, and a method body, say $m$, pruning of the method with respect to the query (i.e. ${\tt Prune}(F(q), m)$) gives a code snippet that is common to both the query and method.  If we consider the similarity score between the query and the pruned code snippet, the score should be an alternative way to quantify the overlap between the query and the method.  We found empirically that if we use this alternative score to rerank the methods retrieved in Phase 1 (i.e. $N_1$), then our ranking of search results improves slightly.  \aroma{} uses the reranked list, which we call $N_2$, in the next phase for clustering and intersection.  Note that the pruning algorithm is greedy, so we may not find the best intersection between two code snippets.  In Section~\ref{sec:microbenchmark}, we show that in very rare cases the greedy pruning algorithm may not give us the best recommended code snippets.

Listing~\ref{lst:pruned} shows a code snippet from the reranked search results for the query code snippet in Listing~\ref{lst:example_code}.  In the code snippet, the highlighted tokens are selected by the pruning algorithm to maximize the similarity score to the query snippet.

% In this example, the containment score of this method is 1.0 (the maximum possible score) because the multi-set union of the features of the highlighted token matches the feature set of the query code snippet perfectly. That is, the containment score identifies that this method completely contains the query code snippet.

%\ksen{We should show two code snippets that are ranked differently in Phase 2 than in Phase 1.  Also pick the two snippets in a way so that they belong to the same cluster.  This will help us to show the result of intersecting the two snippets to get a recommendation.} \fl{Still working on finding such an example. For this particular query all method bodies with top phase 1 score (59) perfectly contain the query so they got 1.0 similarity scores.}

\subsubsection{Phase III: Cluster and Intersect}

In the final phase, \aroma{} prepares recommendations by clustering and intersecting the reranked search results from the previous phase.  Clustering and intersection are computationally expensive.  Therefore, we pick from the list of search results the top $\eta_2 = 100$ methods whose overlap score with the query is above a threshold $\tau_1 = 0.65$, and run the last phase on them.  In the discussion below, we assume that $N_2$ has been modified to contain the top $\eta_2$ search results.

\paragraph{\textbf{Clustering.}}
\aroma{} clusters together method bodies that are similar to each other. The clustering step is necessary to avoid creating redundant recommendations---for each cluster, only one recommendation is generated. Furthermore, the methods in a cluster may contain unnecessary, extraneous code fragments.  An intersection of the code snippets in a cluster helps to create a concise recommendation by getting rid of these unnecessary code fragments.

A cluster contains method bodies that are similar to each other.   Specifically, a cluster must satisfy the following two constraints:
\begin{enumerate}
\item If we intersect the snippets in a cluster, we should get a code snippet that has more code fragments than the query.  This ensures that \aroma{}'s recommendation (which is obtained by intersecting the snippets in the cluster) is an extension to the query snippet.
\item The pruned code snippets in a cluster are similar to each other.  This is because \aroma{} has been designed to perform search that can tolerate some degree of differences between the query and the results.  As such, two code snippets may overlap with different parts of the query.  If two such code snippets are part of a cluster, then their intersection will not contain the query snippet.  Therefore, the recommendation, which is obtained by intersecting all the snippets in a cluster, will not contain any part of the query.  This is undesirable because we want a recommendation that contains the query and some extra new code.
\end{enumerate}
Moreover, \aroma{} does not require the clusters to be disjoint.

Because of these constraints on a cluster, we cannot simply use a textbook clustering algorithm such as k-means, DBSCAN, or Affinity Propagation.  We tried using those clustering algorithms initially (ignoring the constraints) and got poor results.  Therefore, we developed a custom clustering algorithm that takes the constraints into account.  At a high level, the clustering algorithm starts by treating each method body as a separate cluster.  Then, it iteratively merges a cluster with another cluster with single snippet provided that the merged cluster satisfies the cluster constraints and the size of the recommended code snippet from the merged cluster is minimally reduced.  We next formally describe the clustering algorithm.

We use $N_2(i)$ to denote the tree at index $i$ in the list $N_2$.  A cluster is a tuple of indices of the form $(i_1, \ldots, i_k)$, where $i_j < i_{j+1}$ for all $1 \leq j < k$.  A tuple $(i_1, \ldots, i_k)$ denotes a cluster containing the code snippets $N_2(i_1), \ldots, N_2(i_k)$.  We define the commonality score of the tuple $\tau = (i_1, \ldots, i_k)$ as
\[
{\rm\tt cs}(\tau) = |\cap_{1 \leq j \leq k} F(N_2(i_j))|
\]
Similarly, we define the commonality score of the tuple $\tau = (i_1, \ldots, i_k)$ with respect to the query $q$ as
\[
{\rm\tt csq}(\tau) = |\cap_{1 \leq j \leq k} F({\rm\tt Prune}(F(q), N_2(i_j)))|
\]
%\fl{Fixed ordering of two arguments in ${\rm\tt Prune}$.}
We say that a tuple $\tau=(i_1, \ldots, i_k)$ is a \emph{valid tuple} or a \emph{valid cluster} if
\begin{enumerate}
    \item ${\rm\tt l}(\tau) = {\rm\tt cs}(\tau)/{\rm\tt csq}(\tau)$ is greater than some user-defined threshold $\tau_2$ (which is 1.5 in our experiments).  This ensures that after intersecting all the snippets in the cluster, we get a snippet that is at least $\tau_2$ times bigger than the query code snippet.
    \item ${\rm\tt s}(\tau) = {\rm\tt csq}(\tau)/|F(N_2(i_1))|$ is greater than some user-defined threshold $\tau_3$ (which is 0.9 in our experiments).  This requirement ensures that the trees in the cluster are not too similar to each other.  Specifically, it says that the intersection of the pruned snippets in a cluster should be very similar to the first pruned snippet.
\end{enumerate}

\noindent{}The set of valid tuples $\mathcal{C}$ is computed iteratively as follows:
\begin{enumerate}
    \item $\mathcal{C}_1$ is the set $\{(i) \mid 1\leq i \leq |N_2|\ \mbox{\rm and } (i) \ \mbox{\rm is a valid tuple}\}$.
% \item $\mathcal{C}_{\ell+1} = \{ (i_1, \ldots, i_\ell, i) \mid (i_1, \dots, i_\ell) \in \mathcal{C}_{\ell} \ \mbox{\rm and } i_\ell < i \leq |N_2| \ \mbox{\rm and } (i_1, \ldots, i_\ell, i) \ \mbox{\rm is a valid tuple}\} \cup \mathcal{C}_{\ell}$
    \item $\mathcal{C}_{\ell+1} = \mathcal{C}_{\ell} \cup \{ (i_1, \ldots, i_k, i) \mid (i_1, \dots, i_k) \in \mathcal{C}_{\ell} \ \mbox{\rm and } i_k < i \leq |N_2| \ \mbox{\rm and } (i_1, \ldots, i_k, i)$ is a valid tuple and $\forall j \mbox{ if } {i_k < j \leq |N_2| } \mbox { then }{\rm\tt l}((i_1, \ldots, i_k, i)) \geq {\rm\tt l}((i_1, \ldots, i_k, j)) \}$

\end{enumerate}

\aroma{} computes $\mathcal{C}_1$, $\mathcal{C}_2$, \ldots iteratively until it finds an $\ell$ such that $\mathcal{C}_{\ell} = \mathcal{C}_{\ell+1}$. $\mathcal{C} = \mathcal{C}_{\ell}$ is then the set of all clusters.  We developed this custom clustering algorithm because existing popular clustering algorithms such as k-means, DBSCAN and Affinity Propagation all gave poor recommendations. Our clustering algorithm makes use of several similarity metrics (the containment score, the Jaccard similarity of various feature sets), whereas standard clustering algorithms usually depend on a single notion of distance. We found the current best similarity metric and clustering algorithm through trial and error.

After computing all valid tuples, \aroma{} sorts the tuples in ascending order on the first index in each tuple and then in descending order on the length of each tuple. It also drops any tuple $\tau$ from the list if it is similar (i.e. has a Jaccard similarity more than 0.5) to any tuple appearing before $\tau$ in the sorted list.  This ensures that the recommended code snippets are not too similar to each other.  Let $N_3$ be the sorted list of the remaining clusters.

\paragraph{\textbf{Intersection.}}

\aroma{} creates a recommendation by intersecting all the snippets in each cluster.  The intersection algorithm uses the ${\rm\tt Prune}$ function described in Algorithm~\ref{alg:prune} and ensures that the intersection does not discard any code fragment that is part of the query.  Formally, given a tuple $\tau = (i_1, \ldots, i_k)$, ${\rm\tt Intersect}(\tau, q)$ returns a code snippet that is the intersection of the code snippets $N_2(i_1), \ldots, N_2(i_k)$ while ensuring that we retain any code that is similar to $q$.  ${\rm\tt Intersect}((i_1, \ldots, i_k), q)$ is defined recursively as follows:
\begin{itemize}
    \item ${\rm\tt Intersect}((i_1), q)$ = ${\rm\tt Prune}(F(q), N_2(i_1))$.
    \item ${\rm\tt Intersect}((i_1, i_2), q)$ = $ {\rm\tt Prune}(F(N_2(i_2)) \uplus F(q), N_2(i_1))$.
    \item ${\rm\tt Intersect}((i_1, \ldots, i_j, i_{j+1}), q)$ = ${\rm\tt Prune}(F(N_2(i_{j+1})) \cup F(q), {\rm\tt Intersect}((i_1, \ldots, i_j), q))$.
\end{itemize}
% In the rare case of zero intersecting features, we set the similarity score to 0.

% As an example, consider a cluster consisting of Listings~\ref{lst:pruned} and \ref{lst:intersect}. \aroma{} creates a code recommendation for this cluster by running the ${\rm \tt Intersect}$ algorithm iteratively.  Listing~\ref{lst:pruned} is used as the base for intersecting because it is the first method body in the cluster.  First, the algorithm picks nodes in Listing~\ref{lst:pruned} whose features best match the union of feature sets of the query code snippet, Listing~\ref{lst:example_code} and the second method body, Listing~\ref{lst:intersect}.  Then the resulting code, which is the intersection of the first two methods, intersects with the third method, and so on.  This ensures the nodes whose features are common in the cluster to be present, but the nodes whose features only appear in individual methods to be pruned.  The result of the intersection is shown in Listing \ref{lst:recommend}, which is returned as a code recommendation.

%\fl{Michael's comment: Can't you get trees this way that don't pretty print into legal code? E.g., what if you have an if statement without the condition? [In practice we also bring in the lines necessary to make the code look complete. Should we add this detail?]}

In the running example, Listing \ref{lst:pruned} and Listing \ref{lst:intersect} form a cluster. \aroma{} prunes Listing \ref{lst:pruned} with respect to the union set of features of the query code and Listing \ref{lst:intersect} as the intersection between Listing \ref{lst:pruned} and Listing \ref{lst:intersect}. The result of the intersection is shown in Listing \ref{lst:recommend}, which is returned as the recommended code snippet from this cluster.

Finally, \aroma{} picks the top $K$ (where $K=5$ in our implementation) tuples from $N_3$ and returns the intersection of each tuple with the query code snippet as recommendations.

\begin{lstlisting}[label={lst:pruned},caption={A method body containing the query code snippet in Listing~\ref{lst:example_code}. The highlighted text represents tokens selected in the pruning step.\protect\footnotemark}]
if (!(view instanceof EditText)) {
    view.setOnTouchListener(new View.OnTouchListener() {
        public boolean onTouch(View v, MotionEvent event) {
            hideKeyBoard();
            return false;
        }
    });
}
if ((*@\hl{view}@*) instanceof (*@\hl{ViewGroup}@*)) {
    for (int (*@\hl{i}@*) = (*@\hl{0}@*); (*@\hl{i}@*) < (((*@\hl{ViewGroup}@*)) (*@\hl{view}@*)).(*@\hl{getChildCount}@*)(); (*@\hl{i}@*)++) {
        (*@\hl{View}@*) (*@\hl{innerView}@*) = (((*@\hl{ViewGroup}@*)) (*@\hl{view}@*)).(*@\hl{getChildAt}@*)((*@\hl{i}@*));
        setupUIToHideKeyBoardOnTouch(innerView);
    }
}
\end{lstlisting}
\footnotetext{Adapted from \url{https://github.com/arcbit/arcbit-android/blob/master/app/src/main/java/com/arcbit/arcbit/ui/SendFragment.java\#L468}. Accessed in August 2018.}
% https://github.com/arcbit/arcbit-android/blob/master/app/src/main/java/com/arcbit/arcbit/ui/SendFragment.java#L468
% Original score 59
% Pruned score 1.0

\begin{lstlisting}[label={lst:intersect},caption={Another method containing the query code snippet in Listing~\ref{lst:example_code}. The highlighted text represents tokens selected in the pruning step.\protect\footnotemark}]
if (!(view instanceof EditText)) {
    view.setOnTouchListener(new View.OnTouchListener() {
        public boolean onTouch(View v, MotionEvent event) {
            Utils.toggleSoftKeyBoard(LoginActivity.this, true);
            return false;
        }
    });
}
if ((*@\hl{view}@*) instanceof (*@\hl{ViewGroup}@*)) {
    for (int (*@\hl{i}@*) = (*@\hl{0}@*); (*@\hl{i}@*) < (((*@\hl{ViewGroup}@*)) (*@\hl{view}@*)).(*@\hl{getChildCount}@*)(); (*@\hl{i}@*)++) {
        (*@\hl{View}@*) (*@\hl{innerView}@*) = (((*@\hl{ViewGroup}@*)) (*@\hl{view}@*)).(*@\hl{getChildAt}@*)((*@\hl{i}@*));
        setupUI(innerView);
    }
}
\end{lstlisting}
\footnotetext{Adapted from \url{https://github.com/AppLozic/Applozic-Android-Chat-Sample/blob/master/Applozic-Android-AV-Sample/app/src/main/java/com/applozic/mobicomkit/sample/LoginActivity.java\#L171}. Accessed in August 2018.}
% https://github.com/AppLozic/Applozic-Android-Chat-Sample/blob/master/Applozic-Android-AV-Sample/app/src/main/java/com/applozic/mobicomkit/sample/LoginActivity.java#L171
% Original score 59
% Pruned score 1.0

\begin{lstlisting}[label={lst:recommend},caption={A recommended code snippet created by intersecting code in Listing~\ref{lst:pruned} and Listing~\ref{lst:intersect}. Extra lines are highlighted.}]
(*@\hl{if (!(view instanceof EditText)) \{}@*)
    (*@\hl{view.setOnTouchListener(new View.OnTouchListener() \{}@*)
        (*@\hl{public boolean onTouch(View v, MotionEvent event) \{}@*)
            (*@\hl{// your code...}@*)
            (*@\hl{return false;}@*)
        (*@\hl{\}}@*)
    (*@\hl{\});}@*)
(*@\hl{\}}@*)
if (view instanceof ViewGroup) {
    for (int i = 0; i < ((ViewGroup) view).getChildCount(); i++) {
        View innerView = ((ViewGroup) view).getChildAt(i);
        (*@\hl{setupUIToHideKeyBoardOnTouch(innerView);}@*)
    }
}
\end{lstlisting}

%% file: evaluation.tex
\section{Evaluation of Aroma's Code Recommendation Capabilities} \label{sec:evaluation}

% \ksen{Make everything past tense.  For example, we evaluated ..., we created ... etc.}
%In this section we discuss the uniqueness of \aroma{}'s code recommendation results. We first discuss the unique characteristics of the retrieved ``candidate'' code snippets, and then compare \aroma{}'s code recommendation results with existing code search and code recommendation engines. Specifically, we would like to answer the following research questions:
%
%\begin{itemize}
%    \item RQ1: Is \aroma{} recommending useful code snippets?
%    \item RQ2: How do \aroma{} recommendation results compare with existing usage patterns?
%    \item RQ3: How relevant are the candidate results retrieved by \aroma{} for recommendation?
%\end{itemize}

Our goal in this section is to assess how \aroma{} code recommendation can be useful to programmers. To do so, we collected real-world code snippets from Stack Overflow, used them as query snippets, and inspected the code recommendations provided by \aroma{} to understand how they can add value to programmers in various ways.

\subsection{Datasets}

We instantiated \aroma{} on 5,417 GitHub projects where Java is the main language and Android is the project topic.  We ensured the quality of the corpus by picking projects that are not forked from other projects, and have at least 5 stars.  A previous study \cite{Dejavu} shows that duplication exists pervasively on GitHub. To make sure \aroma{} recommendations are created from multiple different code snippets, rather than the same code snippet duplicated in multiple locations, we removed duplicates at project level, file level, and method level.  We do this by taking hashes of these entities and by comparing these hashes.
%\begin{enumerate}
%  \item \emph{Project-level duplication removal}: For each project, we create a MD5 hash by concatenating the contents of each file in a canonical traversal order, ignoring filenames. For projects with identical hashes, we only keep one and remove the others.
%  \item \emph{File-level duplication removal}: For every file, we create a MD5 hash from its content. For files with identical hashes, we only keep one and remove the others.
%  \item \emph{Method-level duplication removal}: For each method, we create a MD5 hash from its simplified parse tree as defined in Section~\ref{sec:algorithm}. For methods with identical hashes we only keep one and remove the others.
%\end{enumerate}
%
%After these steps, the corpus for \aroma{} recommendation contains 2,417,125 methods. Table~\ref{tab:corpus_summary} presents the duplication rate at each level.
After removing duplicates, the corpus contains 2,417,125 methods.

%Table~\ref{tab:corpus_summary} presents the duplication rate at each level.
%
%\begin{table}[!ht] \label{tab:corpus_summary}
%\caption{GitHub code corpus statistics}
%\centering
%\begin{tabular}{@{}llll@{}}
%\toprule
% & Total count & Duplicate count (rate) & Final count \\
%\midrule
%Projects & 5,417 & 6 (0.1\%) & 5,411 \\
%Files & 359,946 & 22,819 (6.3\%) & 337,127 \\
%Methods & 2,801,792 & 384,667 (13\%) & 2,417,125 \\
%\bottomrule
%\end{tabular}
%\end{table}
%The answers on Stack Overflow often contain example code snippets implementing frequently used Android functionalities, and we want to take these code snippets and examine whether \aroma{} can provide meaningful code recommendations from its corpus of real-world open source Android applications.  We downloaded all questions and answers from Stack Overflow as of June 2018 from~\cite{stackoverflowdump}, and
For evaluation, we picked the 500 most popular questions on Stack Overflow with the \emph{android} tag.  From these questions, we only considered the top voted answers.  From each answer, we extracted all Java code snippets containing at least 3 tokens, a method call, and less than 20 lines, excluding comments.  We randomly picked 64 from this set of Java code snippets.  We then used these code snippets to carry out the experimental evaluations in the following two sections.  In these experiments, we found that on average \aroma{} takes 1.6 seconds end-to-end to create recommendations on a 24-core CPU. The median response time is 1.3s and 95\% queries complete in 4 seconds.  A 24-core server was not necessary to achieve reasonable response time: We reran our experiments on a 4-core desktop machine, and the average response time is 2.9 seconds.  We believe this makes \aroma{} suitable for integration into the development environment as a code recommendation tool.

%    \item If a code snippet contains multiple methods, consider each method body as a separate code snippet and apply the rules above.
%\end{enumerate}

%Since there might be several code snippets in a single answer, we get a total of 554 code snippets out of the 500 top answers, from which we randomly pick 54 code snippets for evaluation.

\subsection{Recommendation Performance on Partial Code Snippets}\label{sec:partial}

% \todo{think of a better title}

In this experiment, we manually created partial code snippets by taking the first half of the statements from each of the 64 code snippets. Since each full code snippet from Stack Overflow  represents a popular coding pattern, we wanted to check whether \aroma{} could recommend the missing statements in the code snippet given the partial query code snippet. We always selected the first half of each code snippet to avoid subjective bias. Since we know how the tool works, we would be inclined to pick the lines that we think will produce better results. On average, the query code snippets were 1 to 5 lines and contained 10 to 100 features.

We could not extract partial query code snippets from 14 out of 64 code snippets because they contained a single statement. Single-statement snippets do get recommendations, but since we do not have a ground truth, we cannot judge their quality objectively. For the remaining 50 query code snippets, \aroma{} recommendations fall into the following two categories.

% For the purpose of truncating original queries, we ignore the 9 queries that contain only one line of code. For the 45 queries that have more than one line, we generate a partial query from each of them. For each partial query, one of the authors examine the top 10 recommendations from  Aroma and marks a result as good candidate for completing of query. Another author will double check, if there is conflict, a third author will be invited for judgment and the decision is made by the majority vote.

\subsubsection{Exact Recommendations.} \label{sec:exact-rec}

In 37 cases (74\%), one of the top 5 \aroma{} recommendations matched the original code snippet.
%\ksen{Address Michael's comment one-way or other.}
%\fl{Added top 5. In practice we only show 5 recommendations.}
Example D in Table~\ref{tab:intro-examples} shows a partial query snippet which included the first two statements in a try-catch block of a Stack Overflow code snippet, and \aroma{} recommended the same error handling code as in the original code snippet.
% \ksen{we need to be consistent in terms of query and recommended code pairs in all Listings }

\subsubsection{Alternative Recommendations.}

In the other 13 cases (26\%), none of the \aroma{} recommended code snippets matched the original snippets.  While in each case the recommended snippets did not contain the original usage pattern, they still fall in some of the categories in Table~\ref{tab:recommendation_summary} which we discuss in the next section.  Example E in Table~\ref{tab:intro-examples} shows a partial code snippet which included one of two common ways to send an object with an \texttt{Intent}. Given the statement, \aroma{} did not recommend the other way to serialize an object in the original code snippet, but suggested a customary way to start an activity with an \texttt{Intent} containing a serialized object.

%\begin{table}[!ht]
%\caption{Categories of \aroma{} Recommendations on Partial Code Snippets}
%\centering
%\begin{tabular}{ll}
%\toprule
%Rediscovering the full code snippet & 33 \\
%Alternative usages & 12 \\
%\bottomrule
%\end{tabular}
%\label{tab:partial}
%\end{table}

\subsection{Recommendation Quality on Full Code Snippets}
\label{sec:manual}

%We manually study the quality of \aroma{} recommendations on 54 query code snippets formulated from Stack Overflow. For each query, we examine the recommendations from \aroma{} and classify them into various categories. This process of manual labeling is done jointly by two authors, and is verified and refined by the other two authors. Next we present the categories in detail.

% For each of the 64 query code snippets, we inspected the recommended code snippets
%, manually assessed their quality,
% and classified the recommended snippets into several categories.  Two of the authors did the manual inspection and categorization.  The other two authors verified the results.  Note that a recommended code snippet can fall under multiple categories.
% For each category we present one representative code recommendation below.
% We picked the most useful code recommendation~\ksen{How do you know that some recommended code is useful?}

In this experiment, we used the 64 code snippets as queries to evaluate the quality of \aroma{}'s recommendations. While the experiment in the previous section used \emph{partial} snippets extracted from each of the 64 code snippets, here we used the \emph{full} code snippets. This meant that we could use all 64 snippets instead of just the 50 used in Section~\ref{sec:partial}, as we did not have to filter out single-statement code snippets.

 We manually inspected the recommended code snippets and determined whether they are useful.  We considered a recommended code snippet to be ``useful'' if in a programming scenario where a programmer writes the query code, they would benefit from seeing the related methods or common usage patterns in the code recommendations.  We classified the recommended snippets into several categories by how the recommended code relates to the query snippet.  The classification is subjective because there is no ``ground truth'' on what the recommended code should be, and the actual usefulness depends on how familiar the programmer is with the language and framework.  Nevertheless, we present the categories and some examples in Table~\ref{tab:intro-examples} to demonstrate the variety of code recommendations \aroma{} can provide.  Two of the authors did the manual inspection and categorization, and two other authors verified the results.

\cb{There is no real discussion of these results. Michael and ICSE reviewers commented on this.}

\subsubsection{Configuring Objects}

In this category, the recommended code suggests additional configurations on objects that are already appearing in the query code. Examples include adding callback handlers, and setting additional flags and properties of an existing object. Listings~\ref{lst:intro-bitmap-query}, \ref{lst:intro-bitmap-rec-1} in the introduction, as well as Example A in Table~\ref{tab:intro-examples} shows examples of this category. These recommendations can be helpful to programmers who are unfamiliar with the idiomatic usages of library methods.

\subsubsection{Error Checking and Handling}

In this category, the recommended code adds null checks and other checks before using an object, or adds a try-catch block that guards the original code snippet. Such additional statements are useful reminders to programmers that the program might enter an erroneous state or even crash at runtime if exceptions and corner cases are not carefully handled. Listings~\ref{lst:intro-bitmap-query}, \ref{lst:intro-bitmap-rec-2} in the introduction show an example of this category.

%decoding a bitmap without incurring high memory consumption. However, the recommended code shows that \texttt{OutOfMemoryError} can still happen with large images and it must be properly handled. The recommended code also sets additional flags for \texttt{BitmapFactory}, which falls under the ``Configuring Objects'' category.

\cb{Comment from Michael: the distinction between the next two categories seems fuzzy}

\subsubsection{Post-processing}

The recommended code extends the query code to perform some common operations on the objects or values computed by the query code. For example, recommended code can show API methods that are commonly called.
%suggest potential APIs to call after getting an instance of a system service.
Example B in Table~\ref{tab:intro-examples} shows an example of this category, where the recommendation applies Gaussian blurring on the decoded bitmap image. This pattern is not obligatory but demonstrates a possible effect that can be applied on the original object. This category of recommendations can help programmers discover related methods for achieving certain tasks.

% \ksen{Get rid of Example C from Table~\ref{tab:intro-examples} to save space.}

%where the recommendation applies Gaussian blurring on the decoded bitmap image. This pattern is not obligatory but demonstrates a possible effect that can be applied on the original object.
%Example C in Table~\ref{tab:intro-examples} shows another example where the query code snippet calculates the device acceleration to detect a ``shake'' gesture, and the recommended code shows that it is common to compare the acceleration to a predefined threshold. It also suggests to call APIs to start and stop the shake gesture handling.

% \textbf{Logic on defined variables} Additional logic of defined variables includes new conditions on the variable or new usage of the variable as parameters for other method calls. In this example \ref{lst:logic-defined-variable}, after getting the width and height of the window, Aroma will recommend changing the layout of the window based on comparison between width and height, and the new size of layout will be decided by passing width and height as parameters.

% \begin{lstlisting}[label={lst:logic-defined-variable},caption={Logic on defined variables}]
% Display  display  =  getWindowManager().getDefaultDisplay();
% Point  size  =  new  Point();
% display.getSize(size);
% int  width  =  size.x;
% int  height  =  size.y;
% (*@\hl{if  (height  >  width)  \{}@*)
%     (*@\hl{getWindow().setLayout((int)(width*.9), (int)(height*.8));}@*)
% (*@\hl{\}  else  \{}@*)
%     (*@\hl{getWindow().setLayout((int)(width*.7), (int)(height*.8));}@*)

\subsubsection{Correlated Statements}

The recommended code adds statements that do not affect the original functionalities of the query code, but rather suggests related statements that commonly appear alongside the query code. In Example C in Table~\ref{tab:intro-examples}, the original code moves the cursor to the end of text in an editable text area, where the recommended code also configures the Android Support Action Bar to show the home button and hide the activity title in order to create a more focused view. These statements are not directly related to the text view, but are common in real-world code.

\subsubsection{Unclustered Recommendations}

In rare cases, the query code snippet could match method bodies that are mostly different from each other.  This results in clusters of size 1.  In these cases, \aroma{} performs no intersection and recommends the full method bodies without any pruning.

The number of recommended code snippets for each category is listed in Table~\ref{tab:recommendation_summary}.  For recommendations that belong to multiple categories, we counted them for each of the categories.  We believed the first four categories all can be useful to programmers in different ways, where the unclustered recommendations may not be.  For 59 out of the 64 query code snippets (92\%), \aroma{} generated at least one useful recommended snippet that falls in the first four categories.

% In this manual inspection, we observed that \aroma{} provided useful recommendations for a majority of query code snippets from the Stack Overflow dataset.

\begin{table}[!ht]
\caption{Categories of \aroma{} code recommendations}
\centering
\begin{tabular}{ll}
\toprule
Configuring Objects & 17 \\
Error Checking and Handling & 14 \\
Post-processing & 16 \\
Correlated Statements & 21 \\
Unclustered Recommendations & 5 \\
\bottomrule
\end{tabular}
\label{tab:recommendation_summary}
\end{table}

% \fl{Table is fixed. Also increased sample size to 64.}

\subsection{Comparison with Pattern-Oriented Code Completion} \label{sec:saner}

Pattern-oriented code completion tools~\cite{Nguyen:2009,Nguyen:2012,Mover:2018} could also be used for code recommendation.  For example, \textsc{GraPacc}~\cite{Nguyen:2012} proposed using mined API usage patterns for code completion. To compare \textsc{GraPacc}'s code recommendation capabilities to \aroma{}'s, we took the dataset of 15 Android API usage patterns manually curated from Stack Overflow posts and Android documentation by the authors of \textsc{BigGroum}~\cite{Mover:2018}.  We used \textsc{BigGroum}'s dataset because this tool extends the pattern-mining tool \textsc{Groum} to scale to large corpora with over 1000 repos. While there are more recent ML-based code completion tools, they focus on completing the next token or predicting the correct API method to invoke, which does not directly compare to \aroma{}.
Among the 15 snippets in this dataset, 11 were found in \textsc{BigGroum} mining results. Therefore, if \textsc{GraPacc} is instantiated on the patterns mined by \textsc{BigGroum}, 11 out of the 15 patterns could be recommended by \textsc{GraPacc}.

%was usedas they both provide suggestions of related usage patterns based on an input code snippet. The difference is the latter requires a pool of usage patterns that must be prepared ahead of time. \textsc{GraPacc}~\cite{Nguyen:2012} is a pattern-oriented, context-sensitive code completion tool. Given an input code snippet, \textsc{GraPacc} finds the most relevant patterns in the pool of usage patterns, which are mined as graph-based object usage patterns using a technique called \textsc{GrouMiner}~\cite{Nguyen:2009}. A recent work~\cite{Mover:2018} reported the original \textsc{GrouMiner} cannot scale to a corpus of 70,000 methods, whereas \textsc{BigGroum}, a proposed scalable usage pattern miner, was deployed on a corpus of 1000 open source Android projects, from which 70,000 patterns were found. In the same work~\cite{Mover:2018}, a dataset of 15 Android API usage patterns manually curated from Stack Overflow posts and Android documentation was used for testing how many usage patterns were captured in the mined patterns.

%  Among the 15 known patterns, 11 of them were found in \textsc{BigGroum} mining results. That is to say, if \textsc{GraPacc} is instantiated on the pattern pool mined by \textsc{BigGroum}, 11 out of 15 patterns can be recommended.

In order to evaluate \aroma{}, we followed the same methodology as in Section~\ref{sec:partial} to create a partial query snippet from each of the 15 full patterns, and checked if any of the \aroma{} recommended code snippets contained the full pattern.  For 14 out of 15 patterns, \aroma{} recommended code containing the original usage patterns, i.e. they are \emph{exact recommendations} as defined in Section~\ref{sec:exact-rec}.  An advantage of \aroma{} is that it could recommend code snippets that do not correspond to any previously mined pattern by \textsc{BigGroum}.  Moreover, \aroma{} could recommend code which may not contain any API usage.

%% file: microbenchmark.tex
\section{Evaluation of Search Recall} \label{sec:microbenchmark}

One of the most important and novel phases of the \aroma{}'s code recommendation algorithm is phase II: prune and rerank, which produces the \emph{reranked search results}.  The purpose of this phase is to rank the search results from phase I (i.e. the light-weight search phase) so that any method containing most parts of the query code is ranked higher than a method body containing a smaller part of the query code.  Therefore, if a method contains the entire query code snippet, it should be ranked top in the reranked search result list.  However, in rare cases this property of \aroma{} may not hold due to two reasons: 1) \aroma{}'s pruning algorithm is greedy and approximate due to efficiency reasons, and 2) the kinds of features that we extract may not be sufficient.

To evaluate the recall of the prune and rerank phase, we created a micro-benchmark dataset by extracting partial query code snippets from existing method bodies in the corpus.  On each of these query snippets, \aroma{} should rank the original method body as number 1 in the reranked search result list, or the original method body should be 100\% similar to the first code snippet in the ranked results.  We created two kinds of query code snippets for this micro-benchmark:
\begin{itemize}
    \item \emph{Contiguous code snippets.} We randomly sampled 1000 method bodies with at least 12 lines of code.  From each method body we take the first 5 lines to form a partial query code snippet.
    \item \emph{Non-contiguous code snippets.} We again randomly sampled 1000 method bodies with at least 12 lines of code. From each method body we randomly sample 5 lines to form a partial query code snippet.
\end{itemize}

We first evaluated \aroma{}'s search recall on this dataset.  We employed statistical bootstrapping to minimize sampling bias from the dataset.  Then, we compared it with alternative setups using clone detectors and conventional search techniques.  The results are reported in Table~\ref{tab:mb_compare}.

\begin{table}[!ht]
\caption{Comparison of recall between a clone detector, conventional search techniques, and \aroma{}}
% \ksen{See Michael's comment about alignment issues in this table.}
\centering
\begin{tabular}{@{}llllll@{}}
\toprule
 & \multicolumn{2}{c}{Contiguous} & \phantom{abc} & \multicolumn{2}{c}{Non-contiguous} \\
 \cmidrule{2-3} \cmidrule{5-6}
 & Recall@1 & Recall@100 && Recall@1 & Recall@100 \\
\midrule
\textsc{SCC} & \multicolumn{2}{c}{(12.2\%)} && \multicolumn{2}{c}{(7.7\%)} \\
Keywords Search & 78.3\% & 96.9\% && 93.0\% & 99.9\% \\
Features Search & 78.3\% & 96.8\% && 88.1\% & 98.6\% \\
% \aroma{} w/o Reranking & \textbf{99.1\%} & \textbf{100\%} && 98.1\% & \textbf{100\%} \\
\aroma{} & \textbf{99.1\%} & \textbf{100\%} && \textbf{98.3\%} & \textbf{100\%} \\
\bottomrule
\end{tabular}
\label{tab:mb_compare}
\end{table}

\subsection{Recall in Prune and Rerank Phase}

Recall@$n$ is defined as the percentage of query code snippets for which the original method body is found in the top $n$ methods in the reranked search result list. In addition to Recall@1, we considered Recall@100 because the first 100 methods in the reranked list are used in the clustering phase to create recommended code.

The result in the last row of Table~\ref{tab:mb_compare} shows that \aroma{} is always able to retrieve the original method in the top 100 methods in the reranked search result list. For 99.1\% of contiguous code queries and for 98.3\% of non-contiguous query code snippets, \aroma{} was able to retrieve the original method as the top-ranked result in the reranked search result list.

Listing~\ref{lst:bad-result} demonstrates a rare case where the original method was not retrieved as the top result. Since \aroma{}'s pruning algorithm is greedy (Section~\ref{sec:pruning}), it erroneously decided to pick the statement at line 7, because the statement contains a lot of features which overlap with that of the query code despite the absence of that statement in the query code.  This results in an imperfect similarity score of 0.984. Since there are other code snippets with similar structures which achieves a perfect similarity score of 1.0, the original method was not retrieved at rank 1. Fortunately, this scenario happens rarely enough that it does not affect overall recall.

\begin{lstlisting}[label={lst:bad-result},caption={A method body in the GitHub corpus.\protect\footnotemark Lines 1--5 and 8 were used as query. The pruning algorithm selected the wrong token (as highlighted) that does not belong to the query, resulting in an imperfect similarity score of 0.984.},numbers=left,numberstyle=\tiny,xleftmargin=\parindent]
int result = 0;
int cur;
int count = 0;
do {
    cur = in.readByte() & 0xff;
    result |= (cur & 0x7f) << (count * 7);
    (*@\hl{count}@*)++;
} while (((cur & 0x80) == 0x80) && count < 5);
if ((cur & 0x80) == 0x80) {
    throw new DexException("invalid LEB128 sequence");
}
return result;
\end{lstlisting}
\footnotetext{Adapted from \url{https://github.com/iqiyi/dexSplitter/blob/master/extra/hack_dx/src/com/android/dex/Leb128.java\#L82}. Accessed in August 2018.}

We employed the statistical bootstrapping method and randomly sampled the dataset 30 times, each time from the entire code corpus (i.e. with replacement), in order to test the robustness of our approach and to reduce sampling bias.  We ran the recall test on each of the 30 samples, and calculated a bootstrapped confidence interval using $\hat{\theta} \pm Z_{\alpha/2} \hat{\sigma}/\sqrt{B}$, where $\hat{\theta}$ is the average recall rate, $\hat{\sigma}$ the standard deviation of the recall rates, $\alpha$ the confidence level set to $0.05$, and $B$ the number of resampling $30$.  For contiguous code snippets, the bootstrapped confidence interval is $99.34\% \pm 0.25\%$ for Recall@1, and $99.97\% \pm 0.03\%$ for Recall@100.  For non-contiguous code snippets, the bootstrapped confidence interval is $98.71\% \pm 0.49\%$ for Recall@1, and $99.80\% \pm 0.26\%$ for Recall@100.  These confidence intervals suggest that \aroma{} is able to robustly recall the original method bodies given partial code snippets.

\subsection{Comparison with Clone Detectors and Conventional Search Techniques} \label{sec:scc}

\aroma{}'s search and pruning phases are somewhat related to clone detection and conventional code search. In principle, \aroma{} can use a clone detector or a conventional code search technique to first retrieve a list of methods that contain the query code snippet, and then cluster and intersect the methods to get recommendations. We tested these alternative setups for search recall on the same micro-benchmark dataset.

\subsubsection{Clone Detectors}

\textsc{SourcererCC}~\cite{Sajnani:2016} is a state-of-the-art clone detector that supports Type-3 clone detection. We wanted to compare \aroma{} with \textsc{SourcererCC} to examine whether a current-generation clone detector can be used as the light-weight search phase in \aroma{}.

We instantiated \textsc{SourcererCC} on the same corpus indexed by \aroma{}.  We then used \textsc{SourcererCC} to find clones of the same 1000 contiguous and non-contiguous queries in the micro-benchmark suite.  \textsc{SourcererCC} retrieved all similar methods above a certain similarity threshold, which is 0.7 by default. However, it does not provide any similarity score between two code snippets, so we were unable to rank the retrieved results and report recall at a specific ranking. We could modify \textsc{SourcererCC} to return the similarity scores, but we do not expect the results to change.

\textsc{SourcererCC}'s recall was 12.2\%  and 7.7\% for contiguous and non-contiguous code queries, respectively.
\textsc{SourcererCC} indexes at method-level granularity, and only returns methods whose entire body matches the query code. We also found that in many cases \textsc{SourcererCC} found code snippets \emph{shorter} than the query snippet. While these are Type-3 clones by definition, they are not useful for generating code recommendations in \aroma{}. Extending \textsc{SourcererCC} to return the methods enclosing the clone snippets found would not work, because it does not consider the methods enclosing the target snippets as ``clones'' in the first place. We worked closely with a member of the \textsc{SourcererCC} team, and found that making \textsc{SourcererCC} find all occurrences of an arbitrary code snippet, contiguous and non-contiguous, would require significant reengineering. Therefore, we conclude that current-generation clone detectors may not suit Aroma's requirements for light-weight search.

\subsubsection{Conventional Search Using TF-IDF and Structural Features} \label{par:tfidf}

We implemented a conventional code search technique using classic TF-IDF~\cite{Salton:1986}. Specifically, instead of creating a binary vector in the featurization stage, we created a normalized TF-IDF vector. We then created the sparse index matrix by combining the sparse vectors for every method body. The $(i, j)^{\rm th}$ entry in the matrix is defined as:
\[
{\rm\it tfidf}(i, j) = (1 + \log {\rm\it tf}(i, j)) \cdot \log \frac{J}{{\rm\it df}(i)}
\]
where ${\rm\it tf}(i, j)$ is the count of occurrences of feature $i$ in method $j$, and ${\rm\it df}(i)$ is the number of methods in which feature $i$ exists. $J$ is the total number of methods.  During retrieval, we created a normalized TF-IDF sparse vector from the query code snippet, and then took its dot product with the feature matrix. Since all vectors are normalized, the result contains the cosine similarity between the feature vectors of the query and of every method. We then returned the list of methods ranked by their cosine similarities.

\subsubsection{Conventional Search Using TF-IDF and Keywords}

We implemented another conventional code search technique by simply treating a method body as a bag of words and using the standard TF-IDF technique for retrieval.  To do so, we extracted words instead of structural features from each token, and used the same vectorization technique as in Section \ref{par:tfidf}.

As shown in Table~\ref{tab:mb_compare}, the recall rates of both conventional search techniques are considerably lower than \aroma{}. We observed that in many cases, though the original method was present in the top 100 results, it was not the top result because there are other methods with higher similarity scores due to more overlapping features or keywords. Without pruning, there is no way to determine how well a method contains the query code snippet.  This experiment shows that pruning is essential in order to create a precise ranked list of search results.

%% file: userstudy.tex
\review{This entire section is new.}

\section{\aroma{} in Deployment}
\label{sec:more-languages}

For deployment, we have implemented \aroma{} for three additional languages: Hack, JavaScript and Python. One advantage of \aroma{} is its language-agnostic nature: accommodating a new language requires only implementing a parser that parses code from the target language into a simplified parse tree (as defined in Section~\ref{sec:algorithm}). The rest of the algorithm, including pruning, clustering and intersecting, all work on the generic form of simplified parse trees. This makes \aroma{} suitable for real-world deployment on a codebase that consists of many different programming languages.

The code recommendations for these languages are similar to those for Java, as shown in Table~\ref{tab:intro-examples}. We have also conducted the same recall experiment as described in Section~\ref{sec:microbenchmark}. This ensures that \aroma{} instantiations on different languages all have high retrieval recall rates and thus are capable of creating code recommendations pertinent to a query. The results are shown in Table~\ref{tab:more_languages}.

\begin{table}[!ht]
\caption{\aroma{} recall performance on different languages}
\centering
\begin{tabular}{@{}llllll@{}}
\toprule
 & \multicolumn{2}{c}{Contiguous} & \phantom{abc} & \multicolumn{2}{c}{Non-contiguous} \\
 \cmidrule{2-3} \cmidrule{5-6}
 & Recall@1 & Recall@100 && Recall@1 & Recall@100 \\
\midrule
\aroma{} for Hack & 98.5\% & 100\% && 98.3\% & 99.9\% \\
\aroma{} for JavaScript & 93.9\% & 99.6\% && \multicolumn{2}{c}{not applicable} \\
\aroma{} for Python & 97.5\% & 99.4\% && \multicolumn{2}{c}{not applicable} \\
\bottomrule
\end{tabular}
\label{tab:more_languages}
\end{table}

The recall rates are on par with the original \aroma{} version on the open-source Java corpus. Non-contiguous test samples were not generated for JavaScript or Python for practical reasons: JavaScript code is often embedded with HTML tags, and Python code structure is dependent on indentation levels. Both language features made generating non-contiguous code queries that resemble real-world code queries particularly difficult. Nevertheless, the high recall rates suggest \aroma{}'s algorithm works well across different languages, and \aroma{} is capable of creating useful code recommendations for each language.

\begin{figure}[!ht]
\includegraphics[width=.8\textwidth]{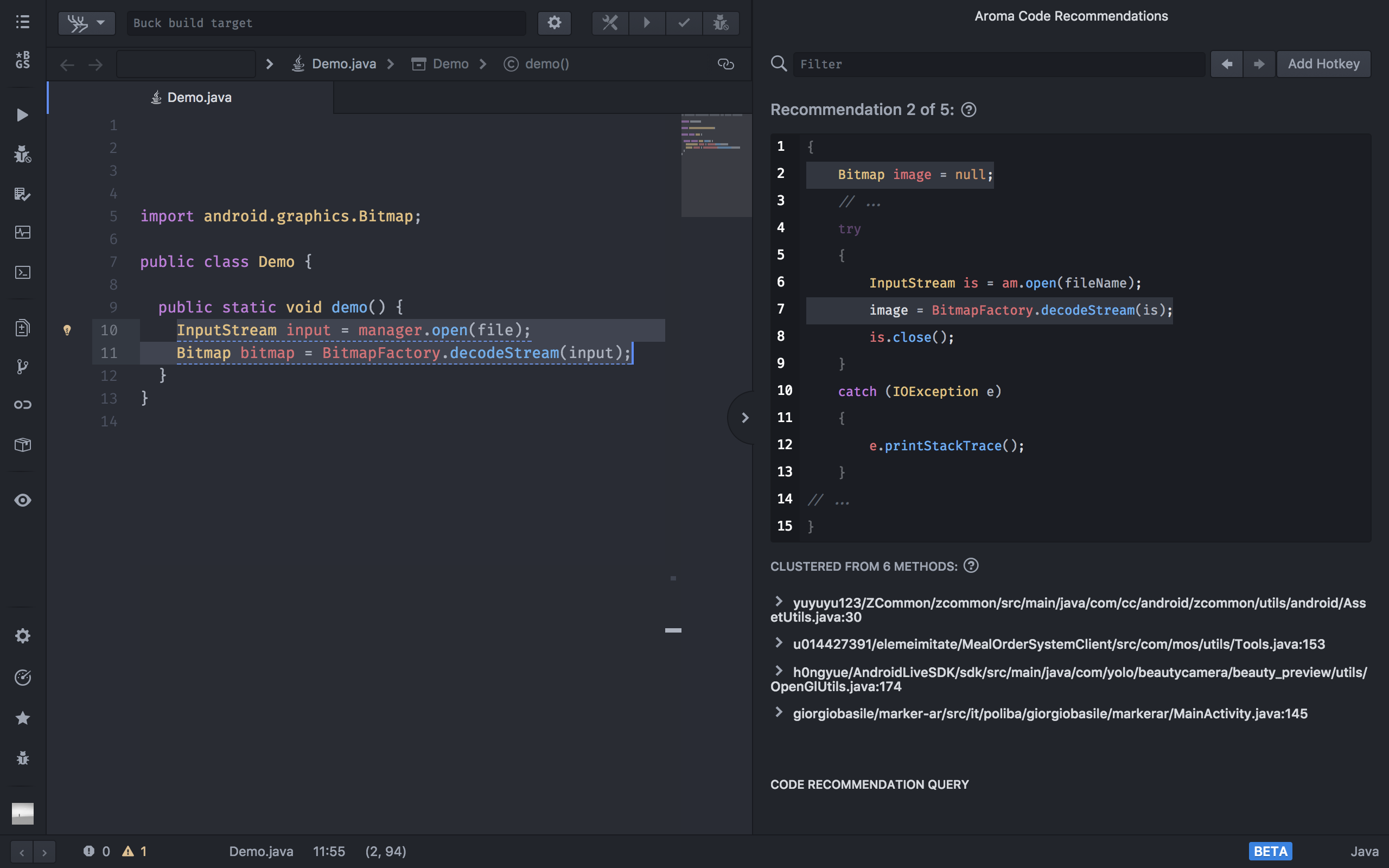}
\caption{\aroma{} code recommendation plugin in an IDE. Recommended code snippets are shown in the side pane for code selected in the main editor.}
\label{fig:aroma_nuclide}
\end{figure}

We have implemented \aroma{} as an IDE plugin for all four languages (Hack, Java, JavaScript and Python). A screenshot of the integrated development environment is shown in Figure~\ref{fig:aroma_nuclide}.

In our setup, we deployed \aroma{} on a dedicated set of servers to respond to queries from all developers. Compared to a distributed setup---where \aroma{} runs on individuals' laptops---this setup allowed for easier control and delivery of new search models. Our search server has 24 cores, and on average took 1.6 seconds end-to-end to create a code recommendation. However, a 24-core server was not necessary to achieve reasonable response time: We reran our experiments in Section~\ref{sec:partial} on a 4-core desktop machine, and the average response time is 2.9 seconds.

The indexing stage (i.e. the creation of the feature vectors, as described in the beginning Section~\ref{sec:reco-algo}) was also designed for scalability. In our setup, we rebuilt the feature vectors for all methods every day. The size of the codebase was comparable to the evaluation dataset. On a 24-core server, building the feature vectors took 20 minutes on average. If \aroma{} is being deployed on a larger-scale codebase, it is possible to implement incremental indexing for only the changed files, without rebuilding the entire feature matrix.

\section{Initial Developer Experience}
\label{sec:userstudy}

We asked 12 Hack programmers to each complete 4 simple programming tasks. For 2 randomly selected tasks, they were allowed to use \aroma{}; for the other 2, they were not. After each programmer completed the tasks, we gave them a brief follow-up survey. For each task, we provided a description of the functionality to achieve, and some incomplete code to start with. The participants were requested to write between 4 to 10 lines of code to implement the desired functionality. Measuring the time taken to complete these tasks with versus without \aroma{} was also initially of interest to us; however, we found that the time taken varied greatly, depending mostly on the experience level of the participant and how familiar they were with the particular frameworks being used in the tasks. It had little correlation with the choice of tools.

We focused on getting some initial feedback on developers' experiences using \aroma{}. The survey began with 2 yes/no questions regarding \aroma{}'s usefulness:

\begin{itemize}
    \item \emph{Did you find \aroma{} useful in completing the programming tasks?} 6 participants answered ``always useful''; 6 other participants answered ``sometimes useful''.
    \item \emph{Did you wish \aroma{} were available in the programming tasks where you were not permitted to use it?} 6 participants answered ``yes''; 4 other participants answered ``sometimes''; 2 participant answered ``no''.
\end{itemize}

 We also asked for more detailed feedback, and found that \aroma{} adds value to existing tools as follows:
\begin{itemize}
    \item \emph{\aroma{} is convenient for discovering usage patterns.} Four participants stated they found \aroma{} is ``convenient'' and they were able to ``quickly find answers''. One participant stated ``being able to see general patterns is nice.'' Another participant commented that when they did not have \aroma{}, they ``used BigGrep\footnote{BigGrep refers to a version of \texttt{grep} that searches the entire codebase.} Code Search to achieve the same goal, but it took longer.''
    \item \emph{\aroma{} is more capable.} One participant commented that \aroma{} is ``more capable at finding results'' with multi-line queries, or queries that do not have exact string matches.
    \item \emph{\aroma{} is as useful as documentation.} Two participants commented that the \aroma{} code recommendations helped them write correct code in the same way as manually curated examples seen in documentations. One said ``it would be a nice backup if there were no documentation,'' another said ``otherwise I would have to read wikis; that would be more tedious.''
\end{itemize}

We also found some common reasons why participants did not feel \aroma{} added any additional value.
\begin{itemize}
    \item \emph{Familiarity with the libraries.} Three participants said they did not find \aroma{} code recommendations to be useful because they either ``had just been working on something like that,'' or ``already knew what to call.'' In these cases they finished the complete code without the help of \aroma{}.
    \item \emph{Simple code search sufficed.} One participant claimed they can get the same amount of information using BigGrep code search. When asked whether clustered results provided additional value, they answered ``the code search results were sufficient for completing the tasks at hand.'' One other participant stated they ``got lucky to find a relevant example using BigGrep, but that might not always be the case.''
\end{itemize}

Based on this initial developer survey, we found that sentiment towards \aroma{} is generally positive, as the participants found \aroma{} useful for conveniently identifying common coding patterns and integrating them into their own code.

%addresses an important challenge that programmers face when writing code: figuring out the common patterns to use library methods, including related methods to call, common errors to check. We also found \aroma{} is a useful addition to the existing coding assistant tools by providing diverse and succinct code recommendations to programmers, based on the code they write.

%% file: related.tex
\section{Related Work} \label{sec:related}

% \ksen{Make sure that we cite any other paper mentioned by Michael Pradel, ICSE reviewers.  Also check if any there is any new related work since our ICSE submission.}
\subsubsection*{Code Search Engines}

Code-to-code search tools like FaCoY~\cite{Kim:2018}and Krugle~\cite{krugle} take a code snippet as query and retrieve relevant code snippets from the corpus. FaCoY aims to find semantically similar results for input queries.  Given a code query, it first searches in a Stack Overflow dataset to find natural language descriptions of the code, and then finds related posts and similar code.  While these code-to-code search tools retrieve similar code at different syntactic and semantic levels, they do not attempt to create concise recommendations from their search results. Further, most of these search engines cannot be instantiated on our code corpus, so we could not experimentally compare \aroma{} with these search engines. For instance, the code search engine FaCoY only provides a VM-based demo that is instantiated on a fixed corpus which is not available publicly. We were also unable to instantiate FaCoY on our corpus for a direct comparison. Most other open-source code search tools, including Krugle and searchcode.com, suffer from the same problem. Instead, we compared \aroma{} with two conventional code search techniques based on featurization and TF-IDF in Section~\ref{sec:scc}, and found that \aroma{}'s pruning-based search technique in Phase II outperforms both techniques.

% These code search engines can only retrieve relevant code snippets but lack of ability of aggregating similar ones, thus they cannot recommend customary patterns as \aroma{} does. Although not suitable for code recommendation, these tools can potentially be served as the search tool for \aroma{} too. Among these three, FaCoY has the best precision for the top 20 results, therefore we also check the possibility of using FaCoY to replace \aroma{}'s search and pruning. Our preliminary results show that FaCoY's processing time is too long its precision after top 20 drops down quickly, therefore it's not a good candidate to perform search for \aroma{}.

Many efforts have been made to improve keyword-based code search~\cite{Sourcerer:2006, Chan:2012, McMillan:2012, Lee:2015, Sachdev:2018}. CodeGenie~\cite{Lemos:2007} uses test cases to search and reuse source code; SNIFF~\cite{Chatterjee:2009} works by inlining API documentation in its code corpus. SNIFF also intersects the search results to provide recommendations, but only targets at resolving natural language queries.  The clustering algorithm in SNIFF is limited and does not take structure into account. Two statements are considered similar if they are syntactically similar after replacing variable names with types. The intersection of two code snippets is the set of statements that appear in both snippets. Due to the strict definition of similarity, SNIFF cannot find large clusters that contain approximately similar code snippets.  Also, SNIFF uses the longest common subsequence algorithm, whose limitations we discuss in Section~\ref{sec:pruning}. MAPO~\cite{Zhong:2009} recommends code examples by mining and indexing associated API usage patterns. Portfolio~\cite{McMillan:2011} retrieves functions and visualizes their usage chains. CodeHow~\cite{Lv:2015} augments the query with API calls which are retrieved from documentation to improve search results. CoCaBu~\cite{Sirres:2018} augments the query with structural code entities. A developer survey~\cite{GoogleCodeSearch} reports the top reason for code search is to find code examples or related APIs, and tools have been created for this need. While these code search techniques focus on creating code examples based on keyword queries, they do not support code-to-code search and recommendation.

\subsubsection*{Clone Detectors}

Clone detectors are designed to detect syntactically identical or highly similar code. \textsc{SourcererCC}~\cite{Sajnani:2016} is a token-based clone detector targeting Type 1, 2, and 3 clones. Compared with other clone detectors that also support Type 3 clones, including NiCad~\cite{Cordy:2011}, Deckard~\cite{Jiang:2007}, and CCFinder ~\cite{Kamiya:2002}, \textsc{SourcererCC} has high precision and recall and also scales to large-size projects. One may repurpose a clone detector to find similar code, but since it is designed for finding highly similar code rather than code that contains the query code snippet---as demonstrated in Section~\ref{sec:scc}---its results are not suitable for code recommendation.

Recent clone detection techniques explored other research directions, from finding semantically similar clones~\cite{Kim:2011, White:2016, Kim:2018, Oreo}, to finding gapped clones~\cite{Ueda:2002} and gapped clones with a large number of edits (large-gapped clones) ~\cite{Wang:2018}. These techniques may excel in finding a particular type of clone, but they sacrifice the precision and recall for Type 1 to 3 clones.

% However, since they are designed for detecting similar codes, the results they retrieved may be too similar to the query and may not have sufficient extra lines to recommend. From Section~\ref{sec:microbenchmark}, we also proved that clone detectors cannot replace the search and pruning mechanism for \aroma{} by comparing with \textsc{SourcererCC} results.

%It achieves scalability by an optimized inverted index to quickly query the potential clones of a given code block and also filtering heuristics based on token ordering to significantly reduce the size of the index, the number of code-block comparisons needed to detect the clones, as well as the number of required token-comparisons needed to judge a potential clone.

\subsubsection*{Pattern Mining and Code Completion}

Code completion can be achieved by different approaches---from extracting the structural context of the code to mining recent histories of editing~\cite{Bruch:2009, Hill:2004, Robbes:2008, Holmes:2005}.  GraPacc~\cite{Nguyen:2012} achieves pattern-oriented code completion by first mining graph-represented coding patterns using GrouMiner~\cite{Nguyen:2009}, then searching for input code to produce code completion suggestions.  More recent work~\cite{Nguyen:2016:APIRec, Nguyen:2016:bytecode, Nguyen:2018} improves code completion by predicting the next API call given a code change.  Pattern-oriented code completion requires mining usage patterns ahead of time, and cannot recommend any code outside of the mined patterns, while \aroma{} does not require pattern mining and recommends snippets on the fly.

%Brunch et. al~\cite{Bruch:2009} propose three code completion algorithms. Hill et. al~\cite{Hill:2004}'s completion of method body relies on code clones. Robbes et. al~\cite{Robbes:2008} improve code completion by recent histories of editing. Strathcona~\cite{Holmes:2005} extracts the structural context of the code under editing and finds its relevant examples.

\subsubsection*{API Documentation Tools}

More techniques exist for improving API documentations and examples.  The work by \citet{Buse:2012} synthesizes API usage examples through data flow analysis, clustering and pattern abstraction.  The work by \citet{Subramanian:2014} augments API documentations with up-to-date source code examples.  MUSE~\cite{Moreno:2015} generates code examples for a specific method using static slicing.  SWIM~\cite{Raghothaman:2016} synthesizes structured call sequences based on a natural language query.  The work by \citet{Robillard:2016} augments API documentation with insights from Stack Overflow.  These tools are limited to API usages and do not generalize to structured code queries.

%\{backup citations\} Chan et. al ~\cite{Chan:2012} search connected API subgraph via text phrases. SWIM ~\cite{Raghothaman:2016} converts user queries into relevant APIs using click–through data from the Bing search engine. It then synthesizes code snippets containing these APIs. MUSE~\cite{Moreno:2015} recommends API usage examples by building index of code examples based on static slicing. Buse et. al~\cite{Buse:2012} synthesize code snippets based on data flow analysis and pattern abstraction. Subramanian et. al~\cite{Subramanian:2014} connect source code to API documentation.

%% file: conclusion.tex
\section{Conclusion}
\label{sec:conclusion}

We presented \aroma{}, a new tool for code recommendation via structural code search. \aroma{} works by first indexing a large code corpus. It takes a code snippet as input, assembles a list of method bodies from the corpus that contain the snippet, and clusters and intersects those method bodies to offer several succinct code recommendations.

To evaluate \aroma{}, we indexed a code corpus with over 2 million Java methods, and performed \aroma{} searches with code snippets chosen from the 500 most popular Stack Overflow questions with the \emph{android} tag. We observed that \aroma{} provided useful recommendations for a majority of these snippets. Moreover, when we used half of the snippet as the query, \aroma{} exactly recommended the second half of the code snippet in 37 out of 50 cases.

Further, we performed a large-scale automated evaluation to test the accuracy of \aroma{} search results. We extracted partial code snippets from existing method bodies in the corpus and performed \aroma{} searches with those snippets as the queries. We found that for 99.1\% of contiguous queries and 98.3\% of non-contiguous queries, \aroma{} retrieved the original method as the top-ranked result. %Using \aroma{} with other programming languages---namely Hack, JavaScript, and Python---yields similar high results.%
We also showed that \aroma{}'s search and pruning algorithms are decidedly better at finding methods \emph{containing} a code snippet than conventional code search techniques.

Finally, we conducted a case study to investigate how programmers interact with \aroma{}, wherein participants completed two short programming tasks with \aroma{} and two without \aroma{}. We found that many participants successfully used \aroma{} to identify common patterns for libraries they were unfamiliar with. In a follow-up survey, a majority of participants stated that they found \aroma{} useful for completing the tasks.

Our ongoing work shows that \aroma{} has the potential to be a powerful developer tool. Though new code is frequently similar to existing code in a repository, currently available code search tools do not leverage this similar code to help programmers add to or improve their code. \aroma{} addresses this problem by identifying common additions or modifications to an input code snippet and presenting them to the programmer in a concise, convenient way.

%% file: aroma_v2.bbl
%%% -*-BibTeX-*-
%%% Do NOT edit. File created by BibTeX with style
%%% ACM-Reference-Format-Journals [18-Jan-2012].

\begin{thebibliography}{43}

%%% ====================================================================
%%% NOTE TO THE USER: you can override these defaults by providing
%%% customized versions of any of these macros before the \bibliography
%%% command.  Each of them MUST provide its own final punctuation,
%%% except for \shownote{}, \showDOI{}, and \showURL{}.  The latter two
%%% do not use final punctuation, in order to avoid confusing it with
%%% the Web address.
%%%
%%% To suppress output of a particular field, define its macro to expand
%%% to an empty string, or better, \unskip, like this:
%%%
%%% \newcommand{\showDOI}[1]{\unskip}   % LaTeX syntax
%%%
%%% \def \showDOI #1{\unskip}           % plain TeX syntax
%%%
%%% ====================================================================

\ifx \showCODEN    \undefined \def \showCODEN     #1{\unskip}     \fi
\ifx \showDOI      \undefined \def \showDOI       #1{#1}\fi
\ifx \showISBNx    \undefined \def \showISBNx     #1{\unskip}     \fi
\ifx \showISBNxiii \undefined \def \showISBNxiii  #1{\unskip}     \fi
\ifx \showISSN     \undefined \def \showISSN      #1{\unskip}     \fi
\ifx \showLCCN     \undefined \def \showLCCN      #1{\unskip}     \fi
\ifx \shownote     \undefined \def \shownote      #1{#1}          \fi
\ifx \showarticletitle \undefined \def \showarticletitle #1{#1}   \fi
\ifx \showURL      \undefined \def \showURL       {\relax}        \fi
% The following commands are used for tagged output and should be
% invisible to TeX
\providecommand\bibfield[2]{#2}
\providecommand\bibinfo[2]{#2}
\providecommand\natexlab[1]{#1}
\providecommand\showeprint[2][]{arXiv:#2}

\bibitem[\protect\citeauthoryear{Bajracharya, Ngo, Linstead, Dou, Rigor, Baldi,
  and Lopes}{Bajracharya et~al\mbox{.}}{2006}]%
        {Sourcerer:2006}
\bibfield{author}{\bibinfo{person}{Sushil Bajracharya}, \bibinfo{person}{Trung
  Ngo}, \bibinfo{person}{Erik Linstead}, \bibinfo{person}{Yimeng Dou},
  \bibinfo{person}{Paul Rigor}, \bibinfo{person}{Pierre Baldi}, {and}
  \bibinfo{person}{Cristina Lopes}.} \bibinfo{year}{2006}\natexlab{}.
\newblock \showarticletitle{Sourcerer: A Search Engine for Open Source Code
  Supporting Structure-based Search}. In \bibinfo{booktitle}{\emph{Companion to
  the 21st ACM SIGPLAN Symposium on Object-oriented Programming Systems,
  Languages, and Applications}} \emph{(\bibinfo{series}{OOPSLA '06})}.
  \bibinfo{publisher}{ACM}, \bibinfo{address}{New York, NY, USA},
  \bibinfo{pages}{681--682}.
\newblock
\showISBNx{1-59593-491-X}
\urldef\tempurl%
\url{https://doi.org/10.1145/1176617.1176671}
\showDOI{\tempurl}


\bibitem[\protect\citeauthoryear{Bruch, Monperrus, and Mezini}{Bruch
  et~al\mbox{.}}{2009}]%
        {Bruch:2009}
\bibfield{author}{\bibinfo{person}{Marcel Bruch}, \bibinfo{person}{Martin
  Monperrus}, {and} \bibinfo{person}{Mira Mezini}.}
  \bibinfo{year}{2009}\natexlab{}.
\newblock \showarticletitle{Learning from Examples to Improve Code Completion
  Systems}. In \bibinfo{booktitle}{\emph{Proceedings of the the 7th Joint
  Meeting of the European Software Engineering Conference and the ACM SIGSOFT
  Symposium on The Foundations of Software Engineering}}
  \emph{(\bibinfo{series}{ESEC/FSE '09})}. \bibinfo{publisher}{ACM},
  \bibinfo{address}{New York, NY, USA}, \bibinfo{pages}{213--222}.
\newblock
\showISBNx{978-1-60558-001-2}
\urldef\tempurl%
\url{https://doi.org/10.1145/1595696.1595728}
\showDOI{\tempurl}


\bibitem[\protect\citeauthoryear{Buse and Weimer}{Buse and Weimer}{2012}]%
        {Buse:2012}
\bibfield{author}{\bibinfo{person}{Raymond P.~L. Buse} {and}
  \bibinfo{person}{Westley Weimer}.} \bibinfo{year}{2012}\natexlab{}.
\newblock \showarticletitle{Synthesizing API Usage Examples}. In
  \bibinfo{booktitle}{\emph{Proceedings of the 34th International Conference on
  Software Engineering}} \emph{(\bibinfo{series}{ICSE '12})}.
  \bibinfo{publisher}{IEEE Press}, \bibinfo{address}{Piscataway, NJ, USA},
  \bibinfo{pages}{782--792}.
\newblock
\showISBNx{978-1-4673-1067-3}
\urldef\tempurl%
\url{http://dl.acm.org/citation.cfm?id=2337223.2337316}
\showURL{%
\tempurl}


\bibitem[\protect\citeauthoryear{Chan, Cheng, and Lo}{Chan
  et~al\mbox{.}}{2012}]%
        {Chan:2012}
\bibfield{author}{\bibinfo{person}{Wing-Kwan Chan}, \bibinfo{person}{Hong
  Cheng}, {and} \bibinfo{person}{David Lo}.} \bibinfo{year}{2012}\natexlab{}.
\newblock \showarticletitle{Searching Connected API Subgraph via Text Phrases}.
  In \bibinfo{booktitle}{\emph{Proceedings of the ACM SIGSOFT 20th
  International Symposium on the Foundations of Software Engineering}}
  \emph{(\bibinfo{series}{FSE '12})}. \bibinfo{publisher}{ACM},
  \bibinfo{address}{New York, NY, USA}, Article \bibinfo{articleno}{10},
  \bibinfo{numpages}{11}~pages.
\newblock
\showISBNx{978-1-4503-1614-9}
\urldef\tempurl%
\url{https://doi.org/10.1145/2393596.2393606}
\showDOI{\tempurl}


\bibitem[\protect\citeauthoryear{Chatterjee, Juvekar, and Sen}{Chatterjee
  et~al\mbox{.}}{2009}]%
        {Chatterjee:2009}
\bibfield{author}{\bibinfo{person}{Shaunak Chatterjee}, \bibinfo{person}{Sudeep
  Juvekar}, {and} \bibinfo{person}{Koushik Sen}.}
  \bibinfo{year}{2009}\natexlab{}.
\newblock \showarticletitle{SNIFF: A Search Engine for Java Using Free-Form
  Queries}. In \bibinfo{booktitle}{\emph{Fundamental Approaches to Software
  Engineering}}, \bibfield{editor}{\bibinfo{person}{Marsha Chechik} {and}
  \bibinfo{person}{Martin Wirsing}} (Eds.). \bibinfo{publisher}{Springer Berlin
  Heidelberg}, \bibinfo{address}{Berlin, Heidelberg},
  \bibinfo{pages}{385--400}.
\newblock
\showISBNx{978-3-642-00593-0}


\bibitem[\protect\citeauthoryear{Cordy and Roy}{Cordy and Roy}{2011}]%
        {Cordy:2011}
\bibfield{author}{\bibinfo{person}{J.~R. Cordy} {and} \bibinfo{person}{C.~K.
  Roy}.} \bibinfo{year}{2011}\natexlab{}.
\newblock \showarticletitle{The NiCad Clone Detector}. In
  \bibinfo{booktitle}{\emph{2011 IEEE 19th International Conference on Program
  Comprehension}}. \bibinfo{pages}{219--220}.
\newblock
\showISSN{1092-8138}
\urldef\tempurl%
\url{https://doi.org/10.1109/ICPC.2011.26}
\showDOI{\tempurl}


\bibitem[\protect\citeauthoryear{Hill and Rideout}{Hill and Rideout}{2004}]%
        {Hill:2004}
\bibfield{author}{\bibinfo{person}{R. Hill} {and} \bibinfo{person}{J.
  Rideout}.} \bibinfo{year}{2004}\natexlab{}.
\newblock \showarticletitle{Automatic method completion}. In
  \bibinfo{booktitle}{\emph{Proceedings. 19th International Conference on
  Automated Software Engineering, 2004.}} \bibinfo{pages}{228--235}.
\newblock
\showISSN{1938-4300}
\urldef\tempurl%
\url{https://doi.org/10.1109/ASE.2004.1342740}
\showDOI{\tempurl}


\bibitem[\protect\citeauthoryear{Holmes and Murphy}{Holmes and Murphy}{2005}]%
        {Holmes:2005}
\bibfield{author}{\bibinfo{person}{R. Holmes} {and} \bibinfo{person}{G.~C.
  Murphy}.} \bibinfo{year}{2005}\natexlab{}.
\newblock \showarticletitle{Using structural context to recommend source code
  examples}. In \bibinfo{booktitle}{\emph{Proceedings. 27th International
  Conference on Software Engineering, 2005. ICSE 2005.}}
  \bibinfo{pages}{117--125}.
\newblock
\showISSN{0270-5257}
\urldef\tempurl%
\url{https://doi.org/10.1109/ICSE.2005.1553554}
\showDOI{\tempurl}


\bibitem[\protect\citeauthoryear{Jiang, Misherghi, Su, and Glondu}{Jiang
  et~al\mbox{.}}{2007}]%
        {Jiang:2007}
\bibfield{author}{\bibinfo{person}{L. Jiang}, \bibinfo{person}{G. Misherghi},
  \bibinfo{person}{Z. Su}, {and} \bibinfo{person}{S. Glondu}.}
  \bibinfo{year}{2007}\natexlab{}.
\newblock \showarticletitle{DECKARD: Scalable and Accurate Tree-Based Detection
  of Code Clones}. In \bibinfo{booktitle}{\emph{29th International Conference
  on Software Engineering (ICSE'07)}}. \bibinfo{pages}{96--105}.
\newblock
\showISSN{0270-5257}
\urldef\tempurl%
\url{https://doi.org/10.1109/ICSE.2007.30}
\showDOI{\tempurl}


\bibitem[\protect\citeauthoryear{Kamiya, Kusumoto, and Inoue}{Kamiya
  et~al\mbox{.}}{2002}]%
        {Kamiya:2002}
\bibfield{author}{\bibinfo{person}{T. Kamiya}, \bibinfo{person}{S. Kusumoto},
  {and} \bibinfo{person}{K. Inoue}.} \bibinfo{year}{2002}\natexlab{}.
\newblock \showarticletitle{CCFinder: a multilinguistic token-based code clone
  detection system for large scale source code}.
\newblock \bibinfo{journal}{\emph{IEEE Transactions on Software Engineering}}
  \bibinfo{volume}{28}, \bibinfo{number}{7} (\bibinfo{date}{July}
  \bibinfo{year}{2002}), \bibinfo{pages}{654--670}.
\newblock
\showISSN{0098-5589}
\urldef\tempurl%
\url{https://doi.org/10.1109/TSE.2002.1019480}
\showDOI{\tempurl}


\bibitem[\protect\citeauthoryear{Kim, Jung, Kim, and Yi}{Kim
  et~al\mbox{.}}{2011}]%
        {Kim:2011}
\bibfield{author}{\bibinfo{person}{H. Kim}, \bibinfo{person}{Y. Jung},
  \bibinfo{person}{S. Kim}, {and} \bibinfo{person}{K. Yi}.}
  \bibinfo{year}{2011}\natexlab{}.
\newblock \showarticletitle{MeCC: memory comparison-based clone detector}. In
  \bibinfo{booktitle}{\emph{2011 33rd International Conference on Software
  Engineering (ICSE)}}. \bibinfo{pages}{301--310}.
\newblock
\showISSN{1558-1225}
\urldef\tempurl%
\url{https://doi.org/10.1145/1985793.1985835}
\showDOI{\tempurl}


\bibitem[\protect\citeauthoryear{Kim, Kim, Bissyand{\'e}, Choi, Li, Klein, and
  Traon}{Kim et~al\mbox{.}}{2018}]%
        {Kim:2018}
\bibfield{author}{\bibinfo{person}{Kisub Kim}, \bibinfo{person}{Dongsun Kim},
  \bibinfo{person}{Tegawend{\'e}~F. Bissyand{\'e}}, \bibinfo{person}{Eunjong
  Choi}, \bibinfo{person}{Li Li}, \bibinfo{person}{Jacques Klein}, {and}
  \bibinfo{person}{Yves~Le Traon}.} \bibinfo{year}{2018}\natexlab{}.
\newblock \showarticletitle{FaCoY: A Code-to-code Search Engine}. In
  \bibinfo{booktitle}{\emph{Proceedings of the 40th International Conference on
  Software Engineering}} \emph{(\bibinfo{series}{ICSE '18})}.
  \bibinfo{publisher}{ACM}, \bibinfo{address}{New York, NY, USA},
  \bibinfo{pages}{946--957}.
\newblock
\showISBNx{978-1-4503-5638-1}
\urldef\tempurl%
\url{https://doi.org/10.1145/3180155.3180187}
\showDOI{\tempurl}


\bibitem[\protect\citeauthoryear{Krugler}{Krugler}{2013}]%
        {krugle}
\bibfield{author}{\bibinfo{person}{Ken Krugler}.}
  \bibinfo{year}{2013}\natexlab{}.
\newblock \bibinfo{booktitle}{\emph{Krugle Code Search Architecture}}.
\newblock \bibinfo{publisher}{Springer New York}, \bibinfo{address}{New York,
  NY}, \bibinfo{pages}{103--120}.
\newblock
\showISBNx{978-1-4614-6596-6}
\urldef\tempurl%
\url{https://doi.org/10.1007/978-1-4614-6596-6_6}
\showDOI{\tempurl}


\bibitem[\protect\citeauthoryear{Lemos, Bajracharya, Ossher, Morla, Masiero,
  Baldi, and Lopes}{Lemos et~al\mbox{.}}{2007}]%
        {Lemos:2007}
\bibfield{author}{\bibinfo{person}{Ot\'{a}vio Augusto~Lazzarini Lemos},
  \bibinfo{person}{Sushil~Krishna Bajracharya}, \bibinfo{person}{Joel Ossher},
  \bibinfo{person}{Ricardo~Santos Morla}, \bibinfo{person}{Paulo~Cesar
  Masiero}, \bibinfo{person}{Pierre Baldi}, {and}
  \bibinfo{person}{Cristina~Videira Lopes}.} \bibinfo{year}{2007}\natexlab{}.
\newblock \showarticletitle{CodeGenie: Using Test-cases to Search and Reuse
  Source Code}. In \bibinfo{booktitle}{\emph{Proceedings of the Twenty-second
  IEEE/ACM International Conference on Automated Software Engineering}}
  \emph{(\bibinfo{series}{ASE '07})}. \bibinfo{publisher}{ACM},
  \bibinfo{address}{New York, NY, USA}, \bibinfo{pages}{525--526}.
\newblock
\showISBNx{978-1-59593-882-4}
\urldef\tempurl%
\url{https://doi.org/10.1145/1321631.1321726}
\showDOI{\tempurl}


\bibitem[\protect\citeauthoryear{Lopes, Maj, Martins, Saini, Yang, Zitny,
  Sajnani, and Vitek}{Lopes et~al\mbox{.}}{2017}]%
        {Dejavu}
\bibfield{author}{\bibinfo{person}{Cristina~V. Lopes}, \bibinfo{person}{Petr
  Maj}, \bibinfo{person}{Pedro Martins}, \bibinfo{person}{Vaibhav Saini},
  \bibinfo{person}{Di Yang}, \bibinfo{person}{Jakub Zitny},
  \bibinfo{person}{Hitesh Sajnani}, {and} \bibinfo{person}{Jan Vitek}.}
  \bibinfo{year}{2017}\natexlab{}.
\newblock \showarticletitle{D{\'e}J\`{a}Vu: A Map of Code Duplicates on
  GitHub}.
\newblock \bibinfo{journal}{\emph{Proc. ACM Program. Lang.}}
  \bibinfo{volume}{1}, \bibinfo{number}{OOPSLA}, Article
  \bibinfo{articleno}{84} (\bibinfo{date}{Oct.} \bibinfo{year}{2017}),
  \bibinfo{numpages}{28}~pages.
\newblock
\showISSN{2475-1421}
\urldef\tempurl%
\url{https://doi.org/10.1145/3133908}
\showDOI{\tempurl}


\bibitem[\protect\citeauthoryear{Lv, Zhang, Lou, Wang, Zhang, and Zhao}{Lv
  et~al\mbox{.}}{2015}]%
        {Lv:2015}
\bibfield{author}{\bibinfo{person}{Fei Lv}, \bibinfo{person}{Hongyu Zhang},
  \bibinfo{person}{Jian-guang Lou}, \bibinfo{person}{Shaowei Wang},
  \bibinfo{person}{Dongmei Zhang}, {and} \bibinfo{person}{Jianjun Zhao}.}
  \bibinfo{year}{2015}\natexlab{}.
\newblock \showarticletitle{CodeHow: Effective Code Search Based on API
  Understanding and Extended Boolean Model (E)}. In
  \bibinfo{booktitle}{\emph{Proceedings of the 2015 30th IEEE/ACM International
  Conference on Automated Software Engineering (ASE)}}
  \emph{(\bibinfo{series}{ASE '15})}. \bibinfo{publisher}{IEEE Computer
  Society}, \bibinfo{address}{Washington, DC, USA}, \bibinfo{pages}{260--270}.
\newblock
\showISBNx{978-1-5090-0025-8}
\urldef\tempurl%
\url{https://doi.org/10.1109/ASE.2015.42}
\showDOI{\tempurl}


\bibitem[\protect\citeauthoryear{Martie, LaToza, and v.~d. Hoek}{Martie
  et~al\mbox{.}}{2015}]%
        {Lee:2015}
\bibfield{author}{\bibinfo{person}{L. Martie}, \bibinfo{person}{T.~D. LaToza},
  {and} \bibinfo{person}{A. v.~d. Hoek}.} \bibinfo{year}{2015}\natexlab{}.
\newblock \showarticletitle{CodeExchange: Supporting Reformulation of
  Internet-Scale Code Queries in Context (T)}. In
  \bibinfo{booktitle}{\emph{2015 30th IEEE/ACM International Conference on
  Automated Software Engineering (ASE)}}. \bibinfo{pages}{24--35}.
\newblock
\urldef\tempurl%
\url{https://doi.org/10.1109/ASE.2015.51}
\showDOI{\tempurl}


\bibitem[\protect\citeauthoryear{McMillan, Grechanik, Poshyvanyk, Fu, and
  Xie}{McMillan et~al\mbox{.}}{2012}]%
        {McMillan:2012}
\bibfield{author}{\bibinfo{person}{C. McMillan}, \bibinfo{person}{M.
  Grechanik}, \bibinfo{person}{D. Poshyvanyk}, \bibinfo{person}{C. Fu}, {and}
  \bibinfo{person}{Q. Xie}.} \bibinfo{year}{2012}\natexlab{}.
\newblock \showarticletitle{Exemplar: A Source Code Search Engine for Finding
  Highly Relevant Applications}.
\newblock \bibinfo{journal}{\emph{IEEE Transactions on Software Engineering}}
  \bibinfo{volume}{38}, \bibinfo{number}{5} (\bibinfo{date}{Sept}
  \bibinfo{year}{2012}), \bibinfo{pages}{1069--1087}.
\newblock
\showISSN{0098-5589}
\urldef\tempurl%
\url{https://doi.org/10.1109/TSE.2011.84}
\showDOI{\tempurl}


\bibitem[\protect\citeauthoryear{McMillan, Grechanik, Poshyvanyk, Xie, and
  Fu}{McMillan et~al\mbox{.}}{2011}]%
        {McMillan:2011}
\bibfield{author}{\bibinfo{person}{Collin McMillan}, \bibinfo{person}{Mark
  Grechanik}, \bibinfo{person}{Denys Poshyvanyk}, \bibinfo{person}{Qing Xie},
  {and} \bibinfo{person}{Chen Fu}.} \bibinfo{year}{2011}\natexlab{}.
\newblock \showarticletitle{Portfolio: Finding Relevant Functions and Their
  Usage}. In \bibinfo{booktitle}{\emph{Proceedings of the 33rd International
  Conference on Software Engineering}} \emph{(\bibinfo{series}{ICSE '11})}.
  \bibinfo{publisher}{ACM}, \bibinfo{address}{New York, NY, USA},
  \bibinfo{pages}{111--120}.
\newblock
\showISBNx{978-1-4503-0445-0}
\urldef\tempurl%
\url{https://doi.org/10.1145/1985793.1985809}
\showDOI{\tempurl}


\bibitem[\protect\citeauthoryear{Moreno, Bavota, Di~Penta, Oliveto, and
  Marcus}{Moreno et~al\mbox{.}}{2015}]%
        {Moreno:2015}
\bibfield{author}{\bibinfo{person}{Laura Moreno}, \bibinfo{person}{Gabriele
  Bavota}, \bibinfo{person}{Massimiliano Di~Penta}, \bibinfo{person}{Rocco
  Oliveto}, {and} \bibinfo{person}{Andrian Marcus}.}
  \bibinfo{year}{2015}\natexlab{}.
\newblock \showarticletitle{How Can I Use This Method?}. In
  \bibinfo{booktitle}{\emph{Proceedings of the 37th International Conference on
  Software Engineering - Volume 1}} \emph{(\bibinfo{series}{ICSE '15})}.
  \bibinfo{publisher}{IEEE Press}, \bibinfo{address}{Piscataway, NJ, USA},
  \bibinfo{pages}{880--890}.
\newblock
\showISBNx{978-1-4799-1934-5}
\urldef\tempurl%
\url{http://dl.acm.org/citation.cfm?id=2818754.2818860}
\showURL{%
\tempurl}


\bibitem[\protect\citeauthoryear{Mover, Sankaranarayanan, Olsen, and
  Chang}{Mover et~al\mbox{.}}{2018}]%
        {Mover:2018}
\bibfield{author}{\bibinfo{person}{S. Mover}, \bibinfo{person}{S.
  Sankaranarayanan}, \bibinfo{person}{R.~B. Olsen}, {and}
  \bibinfo{person}{B.~E. Chang}.} \bibinfo{year}{2018}\natexlab{}.
\newblock \showarticletitle{Mining framework usage graphs from app corpora}. In
  \bibinfo{booktitle}{\emph{2018 IEEE 25th International Conference on Software
  Analysis, Evolution and Reengineering (SANER)}}, Vol.~\bibinfo{volume}{00}.
  \bibinfo{pages}{277--289}.
\newblock
\urldef\tempurl%
\url{https://doi.org/10.1109/SANER.2018.8330216}
\showDOI{\tempurl}


\bibitem[\protect\citeauthoryear{Nguyen, Hilton, Codoban, Nguyen, Mast,
  Rademacher, Nguyen, and Dig}{Nguyen et~al\mbox{.}}{2016a}]%
        {Nguyen:2016:APIRec}
\bibfield{author}{\bibinfo{person}{Anh~Tuan Nguyen}, \bibinfo{person}{Michael
  Hilton}, \bibinfo{person}{Mihai Codoban}, \bibinfo{person}{Hoan~Anh Nguyen},
  \bibinfo{person}{Lily Mast}, \bibinfo{person}{Eli Rademacher},
  \bibinfo{person}{Tien~N. Nguyen}, {and} \bibinfo{person}{Danny Dig}.}
  \bibinfo{year}{2016}\natexlab{a}.
\newblock \showarticletitle{API Code Recommendation Using Statistical Learning
  from Fine-grained Changes}. In \bibinfo{booktitle}{\emph{Proceedings of the
  2016 24th ACM SIGSOFT International Symposium on Foundations of Software
  Engineering}} \emph{(\bibinfo{series}{FSE 2016})}. \bibinfo{publisher}{ACM},
  \bibinfo{address}{New York, NY, USA}, \bibinfo{pages}{511--522}.
\newblock
\showISBNx{978-1-4503-4218-6}
\urldef\tempurl%
\url{https://doi.org/10.1145/2950290.2950333}
\showDOI{\tempurl}


\bibitem[\protect\citeauthoryear{Nguyen, Nguyen, Nguyen, Tamrawi, Nguyen,
  Al-Kofahi, and Nguyen}{Nguyen et~al\mbox{.}}{2012}]%
        {Nguyen:2012}
\bibfield{author}{\bibinfo{person}{Anh~Tuan Nguyen},
  \bibinfo{person}{Tung~Thanh Nguyen}, \bibinfo{person}{Hoan~Anh Nguyen},
  \bibinfo{person}{Ahmed Tamrawi}, \bibinfo{person}{Hung~Viet Nguyen},
  \bibinfo{person}{Jafar Al-Kofahi}, {and} \bibinfo{person}{Tien~N. Nguyen}.}
  \bibinfo{year}{2012}\natexlab{}.
\newblock \showarticletitle{Graph-based Pattern-oriented, Context-sensitive
  Source Code Completion}. In \bibinfo{booktitle}{\emph{Proceedings of the 34th
  International Conference on Software Engineering}}
  \emph{(\bibinfo{series}{ICSE '12})}. \bibinfo{publisher}{IEEE Press},
  \bibinfo{address}{Piscataway, NJ, USA}, \bibinfo{pages}{69--79}.
\newblock
\showISBNx{978-1-4673-1067-3}
\urldef\tempurl%
\url{http://dl.acm.org/citation.cfm?id=2337223.2337232}
\showURL{%
\tempurl}


\bibitem[\protect\citeauthoryear{Nguyen, Tran, Phan, Nguyen, Truong, Nguyen,
  Nguyen, and Nguyen}{Nguyen et~al\mbox{.}}{2018}]%
        {Nguyen:2018}
\bibfield{author}{\bibinfo{person}{Thanh Nguyen}, \bibinfo{person}{Ngoc Tran},
  \bibinfo{person}{Hung Phan}, \bibinfo{person}{Trong Nguyen},
  \bibinfo{person}{Linh Truong}, \bibinfo{person}{Anh~Tuan Nguyen},
  \bibinfo{person}{Hoan~Anh Nguyen}, {and} \bibinfo{person}{Tien~N. Nguyen}.}
  \bibinfo{year}{2018}\natexlab{}.
\newblock \showarticletitle{Complementing Global and Local Contexts in
  Representing API Descriptions to Improve API Retrieval Tasks}. In
  \bibinfo{booktitle}{\emph{Proceedings of the 2018 26th ACM Joint Meeting on
  European Software Engineering Conference and Symposium on the Foundations of
  Software Engineering}} \emph{(\bibinfo{series}{ESEC/FSE 2018})}.
  \bibinfo{publisher}{ACM}, \bibinfo{address}{New York, NY, USA},
  \bibinfo{pages}{551--562}.
\newblock
\showISBNx{978-1-4503-5573-5}
\urldef\tempurl%
\url{https://doi.org/10.1145/3236024.3236036}
\showDOI{\tempurl}


\bibitem[\protect\citeauthoryear{Nguyen, Nguyen, Pham, Al-Kofahi, and
  Nguyen}{Nguyen et~al\mbox{.}}{2009}]%
        {Nguyen:2009}
\bibfield{author}{\bibinfo{person}{Tung~Thanh Nguyen},
  \bibinfo{person}{Hoan~Anh Nguyen}, \bibinfo{person}{Nam~H. Pham},
  \bibinfo{person}{Jafar~M. Al-Kofahi}, {and} \bibinfo{person}{Tien~N.
  Nguyen}.} \bibinfo{year}{2009}\natexlab{}.
\newblock \showarticletitle{Graph-based Mining of Multiple Object Usage
  Patterns}. In \bibinfo{booktitle}{\emph{Proceedings of the the 7th Joint
  Meeting of the European Software Engineering Conference and the ACM SIGSOFT
  Symposium on The Foundations of Software Engineering}}
  \emph{(\bibinfo{series}{ESEC/FSE '09})}. \bibinfo{publisher}{ACM},
  \bibinfo{address}{New York, NY, USA}, \bibinfo{pages}{383--392}.
\newblock
\showISBNx{978-1-60558-001-2}
\urldef\tempurl%
\url{https://doi.org/10.1145/1595696.1595767}
\showDOI{\tempurl}


\bibitem[\protect\citeauthoryear{Nguyen, Pham, Vu, and Nguyen}{Nguyen
  et~al\mbox{.}}{2016b}]%
        {Nguyen:2016:bytecode}
\bibfield{author}{\bibinfo{person}{Tam~The Nguyen}, \bibinfo{person}{Hung~Viet
  Pham}, \bibinfo{person}{Phong~Minh Vu}, {and} \bibinfo{person}{Tung~Thanh
  Nguyen}.} \bibinfo{year}{2016}\natexlab{b}.
\newblock \showarticletitle{Learning API Usages from Bytecode: A Statistical
  Approach}. In \bibinfo{booktitle}{\emph{Proceedings of the 38th International
  Conference on Software Engineering}} \emph{(\bibinfo{series}{ICSE '16})}.
  \bibinfo{publisher}{ACM}, \bibinfo{address}{New York, NY, USA},
  \bibinfo{pages}{416--427}.
\newblock
\showISBNx{978-1-4503-3900-1}
\urldef\tempurl%
\url{https://doi.org/10.1145/2884781.2884873}
\showDOI{\tempurl}


\bibitem[\protect\citeauthoryear{Parr}{Parr}{2013}]%
        {ANTLR}
\bibfield{author}{\bibinfo{person}{Terence Parr}.}
  \bibinfo{year}{2013}\natexlab{}.
\newblock \bibinfo{booktitle}{\emph{The Definitive ANTLR 4 Reference}
  (\bibinfo{edition}{2} ed.)}.
\newblock \bibinfo{publisher}{Pragmatic Bookshelf}.
\newblock
\showISBNx{978-1934356999}


\bibitem[\protect\citeauthoryear{Porter}{Porter}{1997}]%
        {Porter:1997}
\bibfield{author}{\bibinfo{person}{M.~F. Porter}.}
  \bibinfo{year}{1997}\natexlab{}.
\newblock \showarticletitle{Readings in Information Retrieval}.
\newblock \bibinfo{publisher}{Morgan Kaufmann Publishers Inc.},
  \bibinfo{address}{San Francisco, CA, USA}, Chapter An Algorithm for Suffix
  Stripping, \bibinfo{pages}{313--316}.
\newblock
\showISBNx{1-55860-454-5}
\urldef\tempurl%
\url{http://dl.acm.org/citation.cfm?id=275537.275705}
\showURL{%
\tempurl}


\bibitem[\protect\citeauthoryear{Raghothaman, Wei, and Hamadi}{Raghothaman
  et~al\mbox{.}}{2016}]%
        {Raghothaman:2016}
\bibfield{author}{\bibinfo{person}{Mukund Raghothaman}, \bibinfo{person}{Yi
  Wei}, {and} \bibinfo{person}{Youssef Hamadi}.}
  \bibinfo{year}{2016}\natexlab{}.
\newblock \showarticletitle{SWIM: Synthesizing What I Mean: Code Search and
  Idiomatic Snippet Synthesis}. In \bibinfo{booktitle}{\emph{Proceedings of the
  38th International Conference on Software Engineering}}
  \emph{(\bibinfo{series}{ICSE '16})}. \bibinfo{publisher}{ACM},
  \bibinfo{address}{New York, NY, USA}, \bibinfo{pages}{357--367}.
\newblock
\showISBNx{978-1-4503-3900-1}
\urldef\tempurl%
\url{https://doi.org/10.1145/2884781.2884808}
\showDOI{\tempurl}


\bibitem[\protect\citeauthoryear{Robbes and Lanza}{Robbes and Lanza}{2008}]%
        {Robbes:2008}
\bibfield{author}{\bibinfo{person}{R. Robbes} {and} \bibinfo{person}{M.
  Lanza}.} \bibinfo{year}{2008}\natexlab{}.
\newblock \showarticletitle{How Program History Can Improve Code Completion}.
  In \bibinfo{booktitle}{\emph{2008 23rd IEEE/ACM International Conference on
  Automated Software Engineering}}. \bibinfo{pages}{317--326}.
\newblock
\showISSN{1938-4300}
\urldef\tempurl%
\url{https://doi.org/10.1109/ASE.2008.42}
\showDOI{\tempurl}


\bibitem[\protect\citeauthoryear{Sachdev, Li, Luan, Kim, Sen, and
  Chandra}{Sachdev et~al\mbox{.}}{2018}]%
        {Sachdev:2018}
\bibfield{author}{\bibinfo{person}{Saksham Sachdev}, \bibinfo{person}{Hongyu
  Li}, \bibinfo{person}{Sifei Luan}, \bibinfo{person}{Seohyun Kim},
  \bibinfo{person}{Koushik Sen}, {and} \bibinfo{person}{Satish Chandra}.}
  \bibinfo{year}{2018}\natexlab{}.
\newblock \showarticletitle{Retrieval on Source Code: A Neural Code Search}. In
  \bibinfo{booktitle}{\emph{Proceedings of the 2Nd ACM SIGPLAN International
  Workshop on Machine Learning and Programming Languages}}
  \emph{(\bibinfo{series}{MAPL 2018})}. \bibinfo{publisher}{ACM},
  \bibinfo{address}{New York, NY, USA}, \bibinfo{pages}{31--41}.
\newblock
\showISBNx{978-1-4503-5834-7}
\urldef\tempurl%
\url{https://doi.org/10.1145/3211346.3211353}
\showDOI{\tempurl}


\bibitem[\protect\citeauthoryear{Sadowski, Stolee, and Elbaum}{Sadowski
  et~al\mbox{.}}{2015}]%
        {GoogleCodeSearch}
\bibfield{author}{\bibinfo{person}{Caitlin Sadowski},
  \bibinfo{person}{Kathryn~T. Stolee}, {and} \bibinfo{person}{Sebastian
  Elbaum}.} \bibinfo{year}{2015}\natexlab{}.
\newblock \showarticletitle{How Developers Search for Code: A Case Study}. In
  \bibinfo{booktitle}{\emph{Proceedings of the 2015 10th Joint Meeting on
  Foundations of Software Engineering}} \emph{(\bibinfo{series}{ESEC/FSE
  2015})}. \bibinfo{publisher}{ACM}, \bibinfo{address}{New York, NY, USA},
  \bibinfo{pages}{191--201}.
\newblock
\showISBNx{978-1-4503-3675-8}
\urldef\tempurl%
\url{https://doi.org/10.1145/2786805.2786855}
\showDOI{\tempurl}


\bibitem[\protect\citeauthoryear{Saini, Farmahinifarahani, Lu, Baldi, and
  Lopes}{Saini et~al\mbox{.}}{2018}]%
        {Oreo}
\bibfield{author}{\bibinfo{person}{Vaibhav Saini}, \bibinfo{person}{Farima
  Farmahinifarahani}, \bibinfo{person}{Yadong Lu}, \bibinfo{person}{Pierre
  Baldi}, {and} \bibinfo{person}{Cristina~V. Lopes}.}
  \bibinfo{year}{2018}\natexlab{}.
\newblock \showarticletitle{Oreo: Detection of Clones in the Twilight Zone}. In
  \bibinfo{booktitle}{\emph{Proceedings of the 2018 26th ACM Joint Meeting on
  European Software Engineering Conference and Symposium on the Foundations of
  Software Engineering}} \emph{(\bibinfo{series}{ESEC/FSE 2018})}.
  \bibinfo{publisher}{ACM}, \bibinfo{address}{New York, NY, USA},
  \bibinfo{pages}{354--365}.
\newblock
\showISBNx{978-1-4503-5573-5}
\urldef\tempurl%
\url{https://doi.org/10.1145/3236024.3236026}
\showDOI{\tempurl}


\bibitem[\protect\citeauthoryear{Sajnani, Saini, Svajlenko, Roy, and
  Lopes}{Sajnani et~al\mbox{.}}{2016}]%
        {Sajnani:2016}
\bibfield{author}{\bibinfo{person}{Hitesh Sajnani}, \bibinfo{person}{Vaibhav
  Saini}, \bibinfo{person}{Jeffrey Svajlenko}, \bibinfo{person}{Chanchal~K.
  Roy}, {and} \bibinfo{person}{Cristina~V. Lopes}.}
  \bibinfo{year}{2016}\natexlab{}.
\newblock \showarticletitle{SourcererCC: Scaling Code Clone Detection to
  Big-code}. In \bibinfo{booktitle}{\emph{Proceedings of the 38th International
  Conference on Software Engineering}} \emph{(\bibinfo{series}{ICSE '16})}.
  \bibinfo{publisher}{ACM}, \bibinfo{address}{New York, NY, USA},
  \bibinfo{pages}{1157--1168}.
\newblock
\showISBNx{978-1-4503-3900-1}
\urldef\tempurl%
\url{https://doi.org/10.1145/2884781.2884877}
\showDOI{\tempurl}


\bibitem[\protect\citeauthoryear{Salton and McGill}{Salton and McGill}{1986}]%
        {Salton:1986}
\bibfield{author}{\bibinfo{person}{Gerard Salton} {and}
  \bibinfo{person}{Michael~J. McGill}.} \bibinfo{year}{1986}\natexlab{}.
\newblock \bibinfo{booktitle}{\emph{Introduction to Modern Information
  Retrieval}}.
\newblock \bibinfo{publisher}{McGraw-Hill, Inc.}, \bibinfo{address}{New York,
  NY, USA}.
\newblock
\showISBNx{0070544840}


\bibitem[\protect\citeauthoryear{Sirres, Bissyand{\'e}, Kim, Lo, Klein, Kim,
  and Traon}{Sirres et~al\mbox{.}}{2018}]%
        {Sirres:2018}
\bibfield{author}{\bibinfo{person}{Raphael Sirres},
  \bibinfo{person}{Tegawend{\'e}~F. Bissyand{\'e}}, \bibinfo{person}{Dongsun
  Kim}, \bibinfo{person}{David Lo}, \bibinfo{person}{Jacques Klein},
  \bibinfo{person}{Kisub Kim}, {and} \bibinfo{person}{Yves~Le Traon}.}
  \bibinfo{year}{2018}\natexlab{}.
\newblock \showarticletitle{Augmenting and structuring user queries to support
  efficient free-form code search}.
\newblock \bibinfo{journal}{\emph{Empirical Software Engineering}}
  \bibinfo{volume}{23}, \bibinfo{number}{5} (\bibinfo{date}{01 Oct}
  \bibinfo{year}{2018}), \bibinfo{pages}{2622--2654}.
\newblock
\showISSN{1573-7616}
\urldef\tempurl%
\url{https://doi.org/10.1007/s10664-017-9544-y}
\showDOI{\tempurl}


\bibitem[\protect\citeauthoryear{Subramanian, Inozemtseva, and
  Holmes}{Subramanian et~al\mbox{.}}{2014}]%
        {Subramanian:2014}
\bibfield{author}{\bibinfo{person}{Siddharth Subramanian},
  \bibinfo{person}{Laura Inozemtseva}, {and} \bibinfo{person}{Reid Holmes}.}
  \bibinfo{year}{2014}\natexlab{}.
\newblock \showarticletitle{Live API Documentation}. In
  \bibinfo{booktitle}{\emph{Proceedings of the 36th International Conference on
  Software Engineering}} \emph{(\bibinfo{series}{ICSE 2014})}.
  \bibinfo{publisher}{ACM}, \bibinfo{address}{New York, NY, USA},
  \bibinfo{pages}{643--652}.
\newblock
\showISBNx{978-1-4503-2756-5}
\urldef\tempurl%
\url{https://doi.org/10.1145/2568225.2568313}
\showDOI{\tempurl}


\bibitem[\protect\citeauthoryear{Treude and Robillard}{Treude and
  Robillard}{2016}]%
        {Robillard:2016}
\bibfield{author}{\bibinfo{person}{Christoph Treude} {and}
  \bibinfo{person}{Martin~P. Robillard}.} \bibinfo{year}{2016}\natexlab{}.
\newblock \showarticletitle{Augmenting API Documentation with Insights from
  Stack Overflow}. In \bibinfo{booktitle}{\emph{Proceedings of the 38th
  International Conference on Software Engineering}}
  \emph{(\bibinfo{series}{ICSE '16})}. \bibinfo{publisher}{ACM},
  \bibinfo{address}{New York, NY, USA}, \bibinfo{pages}{392--403}.
\newblock
\showISBNx{978-1-4503-3900-1}
\urldef\tempurl%
\url{https://doi.org/10.1145/2884781.2884800}
\showDOI{\tempurl}


\bibitem[\protect\citeauthoryear{Ueda, Kamiya, Kusumoto, and Inoue}{Ueda
  et~al\mbox{.}}{2002}]%
        {Ueda:2002}
\bibfield{author}{\bibinfo{person}{Y. Ueda}, \bibinfo{person}{T. Kamiya},
  \bibinfo{person}{S. Kusumoto}, {and} \bibinfo{person}{K. Inoue}.}
  \bibinfo{year}{2002}\natexlab{}.
\newblock \showarticletitle{On detection of gapped code clones using gap
  locations}. In \bibinfo{booktitle}{\emph{Ninth Asia-Pacific Software
  Engineering Conference, 2002.}} \bibinfo{pages}{327--336}.
\newblock
\showISSN{1530-1362}
\urldef\tempurl%
\url{https://doi.org/10.1109/APSEC.2002.1183002}
\showDOI{\tempurl}


\bibitem[\protect\citeauthoryear{Verlaguet and Menghrajani}{Verlaguet and
  Menghrajani}{2014}]%
        {hacklang}
\bibfield{author}{\bibinfo{person}{Julien Verlaguet} {and}
  \bibinfo{person}{Alok Menghrajani}.} \bibinfo{year}{2014}\natexlab{}.
\newblock \bibinfo{title}{Hack: a new programming language for HHVM}.
\newblock
  \bibinfo{howpublished}{\url{https://code.fb.com/developer-tools/hack-a-new-programming-language-for-hhvm/}}.
\newblock


\bibitem[\protect\citeauthoryear{Wang, Svajlenko, Wu, Xu, and Roy}{Wang
  et~al\mbox{.}}{2018}]%
        {Wang:2018}
\bibfield{author}{\bibinfo{person}{Pengcheng Wang}, \bibinfo{person}{Jeffrey
  Svajlenko}, \bibinfo{person}{Yanzhao Wu}, \bibinfo{person}{Yun Xu}, {and}
  \bibinfo{person}{Chanchal~K. Roy}.} \bibinfo{year}{2018}\natexlab{}.
\newblock \showarticletitle{CCAligner: A Token Based Large-gap Clone Detector}.
  In \bibinfo{booktitle}{\emph{Proceedings of the 40th International Conference
  on Software Engineering}} \emph{(\bibinfo{series}{ICSE '18})}.
  \bibinfo{publisher}{ACM}, \bibinfo{address}{New York, NY, USA},
  \bibinfo{pages}{1066--1077}.
\newblock
\showISBNx{978-1-4503-5638-1}
\urldef\tempurl%
\url{https://doi.org/10.1145/3180155.3180179}
\showDOI{\tempurl}


\bibitem[\protect\citeauthoryear{White, Tufano, Vendome, and Poshyvanyk}{White
  et~al\mbox{.}}{2016}]%
        {White:2016}
\bibfield{author}{\bibinfo{person}{Martin White}, \bibinfo{person}{Michele
  Tufano}, \bibinfo{person}{Christopher Vendome}, {and} \bibinfo{person}{Denys
  Poshyvanyk}.} \bibinfo{year}{2016}\natexlab{}.
\newblock \showarticletitle{Deep Learning Code Fragments for Code Clone
  Detection}. In \bibinfo{booktitle}{\emph{Proceedings of the 31st IEEE/ACM
  International Conference on Automated Software Engineering}}
  \emph{(\bibinfo{series}{ASE 2016})}. \bibinfo{publisher}{ACM},
  \bibinfo{address}{New York, NY, USA}, \bibinfo{pages}{87--98}.
\newblock
\showISBNx{978-1-4503-3845-5}
\urldef\tempurl%
\url{https://doi.org/10.1145/2970276.2970326}
\showDOI{\tempurl}


\bibitem[\protect\citeauthoryear{Zhong, Xie, Zhang, Pei, and Mei}{Zhong
  et~al\mbox{.}}{2009}]%
        {Zhong:2009}
\bibfield{author}{\bibinfo{person}{Hao Zhong}, \bibinfo{person}{Tao Xie},
  \bibinfo{person}{Lu Zhang}, \bibinfo{person}{Jian Pei}, {and}
  \bibinfo{person}{Hong Mei}.} \bibinfo{year}{2009}\natexlab{}.
\newblock \showarticletitle{MAPO: Mining and Recommending API Usage Patterns}.
  In \bibinfo{booktitle}{\emph{ECOOP 2009 -- Object-Oriented Programming}},
  \bibfield{editor}{\bibinfo{person}{Sophia Drossopoulou}} (Ed.).
  \bibinfo{publisher}{Springer Berlin Heidelberg}, \bibinfo{address}{Berlin,
  Heidelberg}, \bibinfo{pages}{318--343}.
\newblock
\showISBNx{978-3-642-03013-0}


\end{thebibliography}
